%% file: iclr2026_conference.tex
\documentclass{article} 
\usepackage{iclr2026_conference,times}

\input{math_commands.tex}

\usepackage{xcolor}
\definecolor{iclrblue}{RGB}{0,0,140} 

\usepackage[
    colorlinks = true,
    linkcolor  = iclrblue,
    citecolor  = iclrblue,
    urlcolor   = iclrblue
]{hyperref}
\usepackage{url}
\usepackage{multirow}
\usepackage{stmaryrd}

\usepackage{booktabs}
\usepackage{multirow}
\usepackage{colortbl}   
\usepackage{xcolor}     
\usepackage{subcaption}

\usepackage[most,skins,breakable]{tcolorbox}
\usepackage{listings}

\definecolor{lightgray}{gray}{0.9}

\usepackage{booktabs,tabularx,url}
\usepackage{siunitx}
\usepackage{booktabs, multirow, makecell, rotating, xcolor, colortbl}
\usepackage[flushleft]{threeparttable}

\newcommand{\rot}[1]{\rotatebox{90}{#1}}
\definecolor{grpA}{HTML}{E8F4FF}     
\definecolor{grpB}{HTML}{FFE8EB}     
\definecolor{grpC}{HTML}{E8FFEF}     
\definecolor{grpD}{HTML}{FFF0E0} 
\definecolor{grpE}{HTML}{FFF0F6} 
\definecolor{avgcol}{gray}{0.9}      
\definecolor{dscol}{gray}{0.95}      
\definecolor{latcol}{RGB}{210,235,255} 
\definecolor{latcpucol}{RGB}{210,235,255} 
\definecolor{latgpucol}{RGB}{170,215,255} 

\usepackage{pifont}
\usepackage{wrapfig}

\title{\flagship{}: Towards Smaller Visual Document Retrievers}

\author{Paul Teiletche$^{1,2}$
\quad Quentin Macé$^{1,3}$ 
\quad Max Conti$^{1}$ 
\quad Antonio Loison$^{1}$ 
\\
\textbf{Gautier Viaud}$^{1}$ 
\quad \textbf{Pierre Colombo}$^{3,4}$
\quad \textbf{Manuel Faysse}$^{1,3}$
\\\\
$^{1}$Illuin Technology 
\quad$^2$EPFL
\quad$^3$CentraleSupélec, Paris-Saclay 
\quad $^4$Equall.ai 
\\
\small \url{paul.teiletche@epfl.ch}}
\definecolor{ownpurple}{RGB}{111, 8, 201}
\definecolor{sportingreen}{RGB}{0, 128, 87}

\iclrfinalcopy 

\ificlrfinal
  \renewcommand{\textcolor}[2]{#2}
\fi

\input{macros}

\begin{document}

\maketitle

\begin{abstract}
    \textcolor{red}{Retrieving specific information from a large corpus of documents is a prevalent industrial use case of modern AI, notably due to the popularity of Retrieval-Augmented Generation (RAG) systems. Although neural document retrieval models have historically operated exclusively in the text space, Visual Document Retrieval (VDR) models -- large vision–language decoders repurposed as embedding models which directly work with page screenshots as inputs -- are increasingly popular due to the performance and indexing latency gains they offer.
    In this work, we show that, while cost-efficient, this approach of repurposing generative models bottlenecks retrieval performance.}
    Through controlled experiments, we \textcolor{red}{revisit the entire training pipeline, and} establish a principled recipe for improving visual document retrieval models. We notably measure the impact of attention masking, image resolution, modality alignment data regimes, and late interaction centered contrastive objectives which emerge as central performance factors.
   Building on these insights, we release \flagship{}, a compact \flagshipparameters{}-parameter vision–language encoder that outperforms recent models up to 10 times larger when fine-tuned on document retrieval tasks, \textcolor{red}{enabling efficient inference on cheap CPU hardware and greatly reducing latency and costs while maintaining strong performance. 
   Models, code and data are available at \repo.
}
\end{abstract}



\input{sections/intro}

\input{sections/methodology}

\input{sections/ablations}

\input{sections/flagship}

\input{sections/related_work}

\input{sections/conclusion}

\section*{Ethics Statement}
\noindent\textbf{Environmental Costs.} Training \colflagship required approximately 2{,}000 H100 GPU-hours in total, which we estimate corresponds to 41\,kg of CO$_2$\footnote{Carbon footprint estimated with \textit{Green Algorithms} \citep{Lannelongue2021GreenAlgorithms}: 
$E = t \times P \times \mathrm{PUE},\; \mathrm{CO_{2}e} = E \times \mathrm{CI}$. 
With $t=2000$ GPUh, $P=0.35$ kW (H100 PCIe), $\mathrm{PUE}=1.3$, and $\mathrm{CI}=45$ gCO$_2$/kWh, 
this gives $E \approx 910$ kWh and $\mathrm{CO_{2}e} \approx 41$ kg.}, based on standard assumptions of GPU power draw, datacenter efficiency, and grid carbon intensity. This estimate follows methodologies such as Green Algorithms \citep{Lannelongue2021GreenAlgorithms} and related analyses of the carbon footprint of machine learning \citep{Strubell2019Energy,Patterson2021Carbon}. Across the entire project, all combined experiments totaled about 18k H100-hours. To mitigate costs and promote sustainable research practices, we release all model checkpoints and training artifacts to facilitate reuse, extension, and reproducibility without necessitating retraining. Additionally, this work shows efficiency gains with smaller models to aim to limit the inference costs of visual retrieval, and consequently reduce the environmental footprint. Our model performs query encoding efficiently on CPUs, keeping inference costs low and reducing barriers to adoption. 

\noindent\textbf{Safety and Bias.} From a safety perspective, our encoder-only retriever poses less risk than generative models: it produces fixed-length embeddings rather than free-form content, reducing avenues for harmful content generation, hallucination, or deceptive outputs; nonetheless, retrieval systems can still propagate biases present in the underlying data, which we address through dataset curation open release.

\noindent\textbf{AI Assistance.} Parts of this paper were prepared with the assistance of an AI-based writing tool used for copy editing and stylistic refinement. All generated text was carefully reviewed, verified, and revised by the authors, who take full responsibility for the accuracy and originality of the final manuscript.

\section*{Reproducibility Statement}
For transparency and to foster future work, we release our training data, model checkpoints (base models and adapters), and the complete codebase under the MIT License, as detailed in the main paper and repository. The supplementary material specifies training configurations for all models (also provided in the corresponding HuggingFace repositories), describes our synthetic data generation process, and reports expanded evaluation results to support exact replication.

\ificlrfinal
\section*{Detailed Contributions}
\textbf{PT} is the first author of the project. Notably, he designed the modality alignement codebase, ran and supervised most large scale experiments across modality alignment and contrastive training, coordinated work, and was key in paper writing. 
\textbf{QM} ran large scale ablations on the data mixtures, contrastive training, and was responsible for the final training run.
\textbf{MC} ran multiple experiments, including investigations into model merging and contributed to paper writing.
\textbf{AL} focused on optimizing the data mixture and the contrastive training codebase. He was notably responsible for the initial cross-modality positive transfer results. 
\textbf{GV} and \textbf{PC} are senior contributors who helped with project framing, grant obtention, industry expertise and paper review.
\textbf{MF} is last author and scientific lead of this work. He initiated and closely supervised the project from beginning to end, wrote the initial version of the multimodal contrastive training framework and greatly contributed to paper writing.

\section*{Acknowledgments}
This work was carried out within the framework of the LIAGORA "LabCom", a joint laboratory supported by the French National Research Agency (ANR) and established between ILLUIN Technology and the MICS laboratory of CentraleSupélec. This work was performed using HPC resources from IDRIS with grant AD011016393. We warmly thank Hippolyte Gisserot-Boukhlef and Nicolas Boizard for sharing the controlled experiments LM checkpoints, Antoine Chaffin for his feedback on the modality alignment codebase and insights on Ettin’s modeling, as well as Andi Marafioti, Orr Zohar, and Miquel Farré for their valuable input and help on gathering the modality alignment dataset.
\fi

\bibliography{iclr2026_conference}
\bibliographystyle{iclr2026_conference}

\newpage
\appendix
\input{sections/appendix}

\end{document}

%% file: math_commands.tex
\usepackage{amsmath,amsfonts,bm}

\def\eqref#1{equation~\ref{#1}}

\def\1{\bm{1}}

\DeclareMathAlphabet{\mathsfit}{\encodingdefault}{\sfdefault}{m}{sl}
\SetMathAlphabet{\mathsfit}{bold}{\encodingdefault}{\sfdefault}{bx}{n}



%% file: macros.tex
\def\flagship{\emph{ModernVBERT}}
\def\flagshipbase{\flagship-base}
\def\flagshipembed{\flagship-embed}
\def\biflagship{\emph{BiModernVBERT}}
\def\colflagship{\emph{ColModernVBERT}}
\def\flagshipparameters{250M}

\newcommand{\cmark}{\ding{51}}
\newcommand{\xmark}{\ding{55}}

\def\natcap{\textit{NatCap}}

\def\repo{\url{https://huggingface.com/XXX}}
\ificlrfinal
\def\repo{\url{https://huggingface.co/ModernVBERT}}
\fi

\def\ghrepo{\url{https://github.com/XXX}}
\ificlrfinal
\def\ghrepo{\url{https://github.com/illuin-tech/modernvbert}}
\fi

%% file: sections/intro.tex
\section{\textcolor{red}{Introduction}}\label{sec:intro}

\begin{wrapfigure}[19]{r}{0.525\textwidth} 
  \vspace{-50pt}                          
  \centering
  \includegraphics[width=0.88\linewidth]{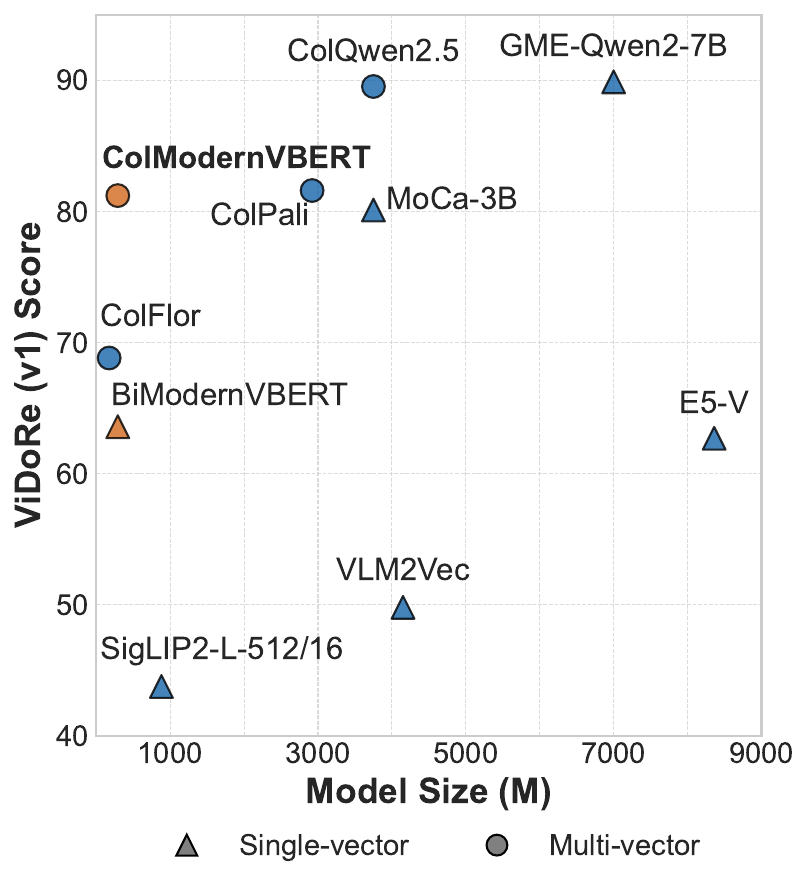}
  \caption{\textbf{Pareto efficiency.} \colflagship{} outperforms models in its category on ViDoRe, achieving a leading performance-size tradeoff.
  }
  \label{fig:pareto_vidore}
\end{wrapfigure}


The ability to quickly locate specific information in vast document collections is a core building block of digital systems today, supporting use cases that range from web search and virtual assistants to enterprise knowledge management. Neural information retrieval (IR) models, and in particular dense retrievers, have become the de facto backbone of modern search systems thanks to their strong semantic matching capabilities and good scalability properties \citep{reimers_sentence-bert_2019, karpukhin_dense_2020, wang_text_2022}.  

This trend is amplified by the widespread adoption of Retrieval-Augmented Generation (RAG) \citep{lewis_retrieval-augmented_2020}, where a retriever is used to select a small set of relevant documents that condition a downstream generator. In such systems, the first-stage retrieval module is a well-known bottleneck: its recall directly upper-bounds the quality of the generated answers, while its latency and indexing costs partially drive the overall system efficiency \citep{lin-byrne-2022-retrieval}. As a result, improving document retrieval, especially for long, complex files such as PDFs, scientific articles, and reports, is a key lever for making industrial RAG deployments more accurate and cost-effective.
\textbf{Visual Document Retrieval.} Historically, document retrieval in these settings has operated purely in the text space. To index PDFs or scans, practitioners first run heavy preprocessing pipelines that include Optical Character Recognition (OCR), layout analysis, and heuristic passage segmentation, before embedding the resulting text spans with a neural encoder. 
This approach suffers from several limitations: OCR and layout parsing can be brittle and slow, complex visual elements such as tables, figures, and typography are often poorly captured, and any error or bias introduced during preprocessing is propagated to the retriever. 

\textit{Visual Document Retrieval} (VDR) has emerged as a compelling alternative to such text-based systems. Rather than indexing pre-extracted textual content, VDR models directly operate on page screenshots: given a user query, they retrieve relevant document pages by matching the query against image-based representations of the pages \citep{faysse2025colpaliefficientdocumentretrieval}. By bypassing OCR and layout parsing, VDR yields simpler end-to-end pipelines, significantly reduces indexing latency, and better exploits visual cues such as layout, figures, and fonts, while achieving strong performance on visually rich benchmarks like ViDoRe. 

\textbf{Limits of Generative VLM Repurposing.} Most current VDR systems are obtained by repurposing large generative vision–language decoders ~\citep{alayrac2022flamingovisuallanguagemodel} as retrieval encoders via post-hoc contrastive fine-tuning \citep{ma2024unifyingmultimodalretrievaldocument,faysse2025colpaliefficientdocumentretrieval,jiang2025vlm2vectrainingvisionlanguagemodels}. While cost-efficient, this design choice bottlenecks retrieval performance and efficiency: model sizes, attention patterns, image resolutions, and training objectives are designed for generative use cases rather than optimized for retrieval which has been shown in text models to be suboptimal \citep{lee2025nvembedimprovedtechniquestraining, gisserotboukhlef2025pretrainencodersmaskedlanguage}. Furthermore, scaling trends \citep{wei2022emergentabilitieslargelanguage} are less pronounced for embedding models; while correlated with model size, strong retrieval performance remains attainable with small models \citep{clavié2024bettermonolingualjapaneseretrievers}.

Recent papers and model releases in the visual retrieval space have claimed performance improvements by scaling the amount of contrastive data and the compute budget \citep{zhang2025gmeimprovinguniversalmultimodal, xu2025llamanemoretrievercolembedtopperforming}, modifying the attention mask \citep{chen2025mocamodalityawarecontinualpretraining}, increasing image resolutions \citep{cohere_introducing_2024} or by introducing more diverse tasks and data sources \citep{jiang2025vlm2vectrainingvisionlanguagemodels}. 


In this work, we attempt to centralize these efforts and systematically disentangle the impact of core design decisions in visual retriever training. 
Through controlled experiments—ranging from language model pretraining to multi-stage, domain-specific fine-tuning, we aim to answer a central question:

\begin{center}
\textit{Which design choices best boost performance in modern visual document retrievers?}
\end{center}

\noindent\textbf{Contribution 1.}\; We revisit core assumptions in visual retriever design, showing that token-level training objectives benefit retrievers by strengthening image–text token alignment—rather than merely producing stronger image embeddings.
Our results indicate that causal attention is suboptimal in document retrieval, with bidirectional masking offering clear improvements in multi-vector settings, and that other parameters such as image resolution data mixes should not be overlooked in the training pipeline.

\noindent\textbf{Contribution 2: \flagship{}.}\; Building on these insights, we release \flagship{}, a small \flagshipparameters{} multimodal encoder that aligns a pretrained language encoder with a vision encoder through Masked Language Modeling (MLM) objective, and \colflagship{} a variant fine-tuned for document retrieval. Despite its modest size and limited training budget, \colflagship{} matches models 10x larger on standard visual document retrieval benchmarks, demonstrating the interest of designing a retrieval focused model from the ground up. 
\textcolor{red}{We release the model, intermediate checkpoints, and the training code at \repo.}

%% file: sections/methodology.tex
\section{Methodology}\label{sec:methodology}



Our analysis aims at quantifying the impact of design decisions made when training visual retrievers. In opposition to previous work, we begin our analysis as early as language model modality alignment and iteratively study design choices by modifying design choices independently to reduce confounding factors as much as possible \citep{allenzhu2025physicslanguagemodels1}.

\noindent\textbf{Controlled Experimental Setup.}
A central point of interest is the impact of causal and bidirectional attention masks. While recently studied for textual representation applications \citep{gisserotboukhlef2025pretrainencodersmaskedlanguage, weller2025seqvsseqopen}, we extend the experiment to the vision modality. We use checkpoints released by \citet{gisserotboukhlef2025pretrainencodersmaskedlanguage} which consist in a series of identical 210M parameter transformer models based on the Llama architecture \citep{touvron_llama_2023} trained on 100B tokens that differ only in their attention masking strategy during language model training but that are perfectly identical in terms of training data seen, model size and architecture, learning rate scheduling, etc... The checkpoints we use are \texttt{enc} a bidirectional encoder trained with Masked Language Modeling (MLM), \texttt{dec}, a causal decoder trained with next token prediction, and \texttt{dec-enc} a causal decoder annealed over the end of its textual training by removing the causal mask and switching the training objective to MLM. For the vision tower, we employ the vision component of \texttt{siglip2-base-16b-512} \citep{tschannen2025siglip2multilingualvisionlanguage}, a 86M parameter vision transformer contrastively trained on billions of text-image pairs.
All ablations thus stem from iso-data controlled setups, and as further described, are further trained on the same data sequence, with the same batch sizes, optimizers, schedulers and on the same hardware.

\begin{figure}[t]
    \centering
    \includegraphics[width=0.99\textwidth]{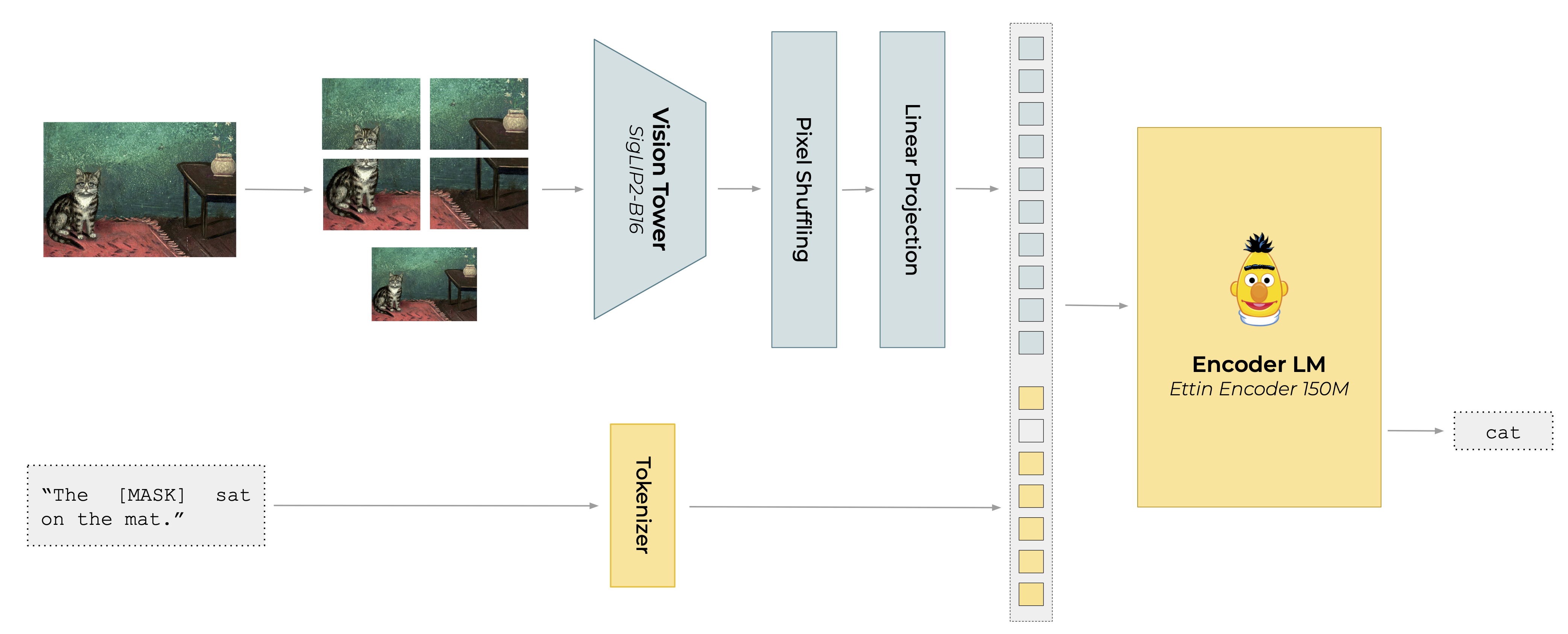}
    \caption{
    \textbf{MLM-based early fusion architecture.} The visual encoder produces patch representations, which are passed to a language model. 
    Our end-to-end bidirectional attention fused architecture is trained with Masked Language Modeling objectives and is perfectly suited for sequence and token-level representation tasks.}
    \label{fig:lf_architecture}
\end{figure}

\noindent\textbf{Model Architecture.}  
Our analysis are not centered around model architectures and to draw broadly applicable insights, we design vision-language models following current standard training practices.  In line with most recent work, we employ the early fusion architecture \citep{alayrac2022flamingovisuallanguagemodel} illustrated in Figure~\ref{fig:lf_architecture}, in which visual patch embeddings produced by the vision encoder are projected into the language model input embedding space and concatenated with text token embeddings to encourage joint processing ~\citep{li2022blipbootstrappinglanguageimagepretraining,alayrac2022flamingovisuallanguagemodel,wang2024qwen2vlenhancingvisionlanguagemodels,yang2025qwen251mtechnicalreport,marafioti2025smolvlmredefiningsmallefficient}. As described in \autoref{sec:modality-align}, we generalize the training loss to function both with causal and masked language modeling objectives.  
To handle dynamic resolutions, we split large images into 512$\times$512 pixel tiles as expected by the SigLIP encoder\textcolor{red}{\footnote{\textcolor{red}{Images are downscaled (or upscaled) so that the lengths and widths reach a multiple of 512 pixels to preserve the aspect ratio, padding is used on the smaller side when necessary (i.e. a 1024x1000 px image would be padded to 1024x1024 px).}}}. Following current standard practices, we further process a downscaled version of the full image to improve inter-tile consistency and global visual understanding~\citep{lin2023sphinxjointmixingweights,ye2023ureaderuniversalocrfreevisuallysituated}. \textcolor{red}{The vision tower produces 1024 pixel patch representations for each tile\footnote{\textcolor{red}{The SigLIP tower takes 512x512 px images and process them by 16x16 px patches \citep{dosovitskiy_image_2020}. This results in $(512/16)^2=1024$ patches.}}, which we  compress to 64 tokens through \textit{pixel shuffling}~\citep{shi2016realtimesingleimagevideo}} with a compression ratio $r=4$, following prior work on models of comparable size~\citep{marafioti2025smolvlmredefiningsmallefficient}. \textcolor{red}{We highlight the impact of image resolution and this parameter on the number of visual tokens in Appendix~\ref{appendix:visual_tokens}}.

\noindent\textbf{Training Procedure.} Our experiments focus on retrieval performance. We employ a standard biphasic training procedure, in which we first run modality alignment to train a pretrained textual language model to understand visual inputs through language modeling objectives  ~\citep{liu2023visualinstructiontuning} (\autoref{sec:modality-align}), then rely on a second text-image contrastive learning phase to learn efficient image representations ~\citep{radford2021learningtransferablevisualmodels} (\autoref{sec:contrastive_posttraining}). We further describe the general setup, and detail specific modifications to the default training procedure in the experiment section.

\subsection{Modality Alignment}
\label{sec:modality-align}
We align the vision encoder tower with the language model by training the image embedding projection layer to map visual features into the language model embedding space. The pretrained language model is also fine-tuned with Low-Rank Adapters (LoRA) \citep{hu2021loralowrankadaptationlarge}, allowing both image and text models to adapt jointly while reducing the risk of monomodal performance collapse \citep{alayrac2022flamingovisuallanguagemodel,liu2023visualinstructiontuning,laurençon2024mattersbuildingvisionlanguagemodels,mckinzie2024mm1methodsanalysis,marafioti2025smolvlmredefiningsmallefficient}.

\noindent\textbf{Alignment Loss.} 
For decoder‑based models, we train with Causal Language Modeling (CLM) loss on the text tokens, as standardly done in VLM modality alignment:
\begin{equation}
\mathcal{L}_{\text{CLM}}
  = -\sum_{t=1}^{T} \log P_{\theta}\!\bigl(x_t \mid x_{<t}\bigr),
\end{equation}

where $x_{<t}$ denotes all tokens preceding position $t$. We generalize this training scheme to bidirectional encoders models, by using the Masked Language Modeling (MLM) loss on the textual tokens:
\begin{equation}
\mathcal{L}_{\text{MLM}}
  = -\sum_{t \in \mathcal{M}} \log P_{\theta}\!\bigl(x_t \mid x_{\backslash\mathcal{M}}\bigr),
\end{equation}
where $\mathcal{M}$ is the set of masked token positions and
$x_{\backslash\mathcal{M}}$ is the input with those tokens masked out.

\noindent\textbf{Modality Alignment Corpus.} Models are modality aligned on a large corpus in large parts derived from The Cauldron 2~\citep{laurençon2024mattersbuildingvisionlanguagemodels} and Docmatix~\citep{laurençon2024building}. Our objective being to train document focused retrieval models, we use an adjusted training mixture that upsamples images containing text and documents with varying level of complexities. Our final training corpus consists of approximately 2B text tokens, and includes diverse sources such as web pages, books, and scientific papers. Mixture details are given in Appendix~\ref{appendix:alignment_mixture}. We note that controlling the exact data distribution during this phase enables the models we train to specialize early and achieve good document focused downstream performances which many large models struggle with \citep{liu_improved_2023}.

\noindent\textbf{Parameters.} All models are trained using a masking ratio of 0.5 and user-prompt masking to avoid overfitting on chat-template format~\citep{huertaenochian2024instructionfinetuningdoesprompt,shi2024instructiontuninglossinstructions,allal2025smollm2smolgoesbig}. We employ WSD scheduler~\citep{hu2024minicpmunveilingpotentialsmall} with the first 5\% of the training as warmup, the last 20\% as decay and a maximum learning rate of 1e-4. The ablation models are aligned on 3.5B tokens. We provide additional details on the training setup in Appendix~\ref{appendix:implementation_details}.

\subsection{Contrastive Post-Training}
\label{sec:contrastive_posttraining}
Once the language model has learned to process image tokens jointly with text tokens, we specialize models through a contrastive post-training stage designed to enhance the semantic representation of the output embeddings produced by the model \citep{reimers_sentence-bert_2019}. 



\noindent\textbf{Post-training Pairs.} The post-training dataset used as starting point in our ablations comprises 118\textcolor{red}{k} document-query pairs from the ColPali corpus~\cite{faysse2025colpaliefficientdocumentretrieval} as well as another 118\textcolor{red}{k} of natural image-description pairs from the MSCOCO train set~\citep{lin2015microsoftcococommonobjects}.

\textbf{Contrastive Loss.} We employ the InfoNCE loss~\citep{oord2019representationlearningcontrastivepredictive}, defined as

\begin{equation}
\mathcal{L}_{\text{InfoNCE}}(\mathbf{q},\mathbf{d^+})
  =-\log\frac{\Phi(\mathbf{q},\mathbf{d^+})}{\Phi(\mathbf{q},\mathbf{d^+}) + \sum_{\mathbf{d^-}\in\mathcal{N}_q}\Phi(\mathbf{q},\mathbf{d^-})},
\end{equation}

where $\mathbf{d^+}$ denotes the positive target for the query $\mathbf{q}$, $\mathcal{N}_\mathbf{q}=\mathcal{N}_\mathbf{q}^{\text{in}} \cup \mathcal{N}_\mathbf{q}^{\text{hard}}$ the set of negative targets (in-batch and hard negatives when mentioned), and $\Phi(\mathbf{q},\mathbf{d})$ a similarity function between the token(s) of the query and the documents.\footnote{We use the last (EOS) token for causal models, and mean pool all sequence tokens for bidirectional encoders for single-vector models. Alternatively, we use all document and query tokens without pooling for late interaction matching \citep{faysse2025colpaliefficientdocumentretrieval}. Details in Appendix~\ref{appendix:similarity_functions}}. 
For general-domain post-training we compute the loss symmetrically~\citep{radford2021learningtransferablevisualmodels}.

\textbf{Batches Curation.} In contrastive learning, batch diversity critically impacts retrieval entropy. Overly heterogeneous batches lead to trivial retrievals, while curated batches yield richer training signals. We employ task-aware batching~\citep{li2023generaltextembeddingsmultistage}, grouping documents by source to ensure a homogeneous batch composition.

\subsection{Ablation Evaluation Setup}

The contrastively trained models are evaluated on retrieval and zero-shot classification tasks across multiple domains. Although the main focus remains document retrieval capabilities, evaluated by aggregating scores from the ViDoRe and ViDoRe v2\footnote{\label{fn:vidorev2}We report only the English splits of ViDoRe v2, as our base models are trained on English data only.} \citep{macé2025vidorebenchmarkv2raising} benchmarks (nDCG@5), we also assess more generalist image retrieval capabilities by selecting tasks from MIEB~\citep{xiao2025miebmassiveimageembedding}. For natural image retrieval, we aggregate MSCOCO retrieval ~\citep{lin2015microsoftcococommonobjects} and Flickr30k retrieval (nDCG@10) \citep{plummer2016flickr30kentitiescollectingregiontophrase} test sets. Finally, following practices in \citep{muennighoff_mteb_2022}, we assess both zero-shot and fine-tuning abilities of our models on general classification tasks. Specifically, we measure classification accuracy by fine-tuning a logistic regression head on top of our model's embedding on Stanford Cars~\citep{6755945} and Food101~\citep{10.1007/978-3-319-10599-4_29}, and we evaluate zero-shot performance on FER2013~\citep{khaireddin2021facialemotionrecognitionstate} and EuroSAT~\citep{helber2019eurosatnoveldatasetdeep} and aggregate the results.

%% file: sections/ablations.tex
\section{What Makes a Great Visual Retriever?}

\begin{wrapfigure}[19]{r}{0.5\textwidth} 
  \vspace{4pt}                          
  \includegraphics[width=0.95\linewidth]{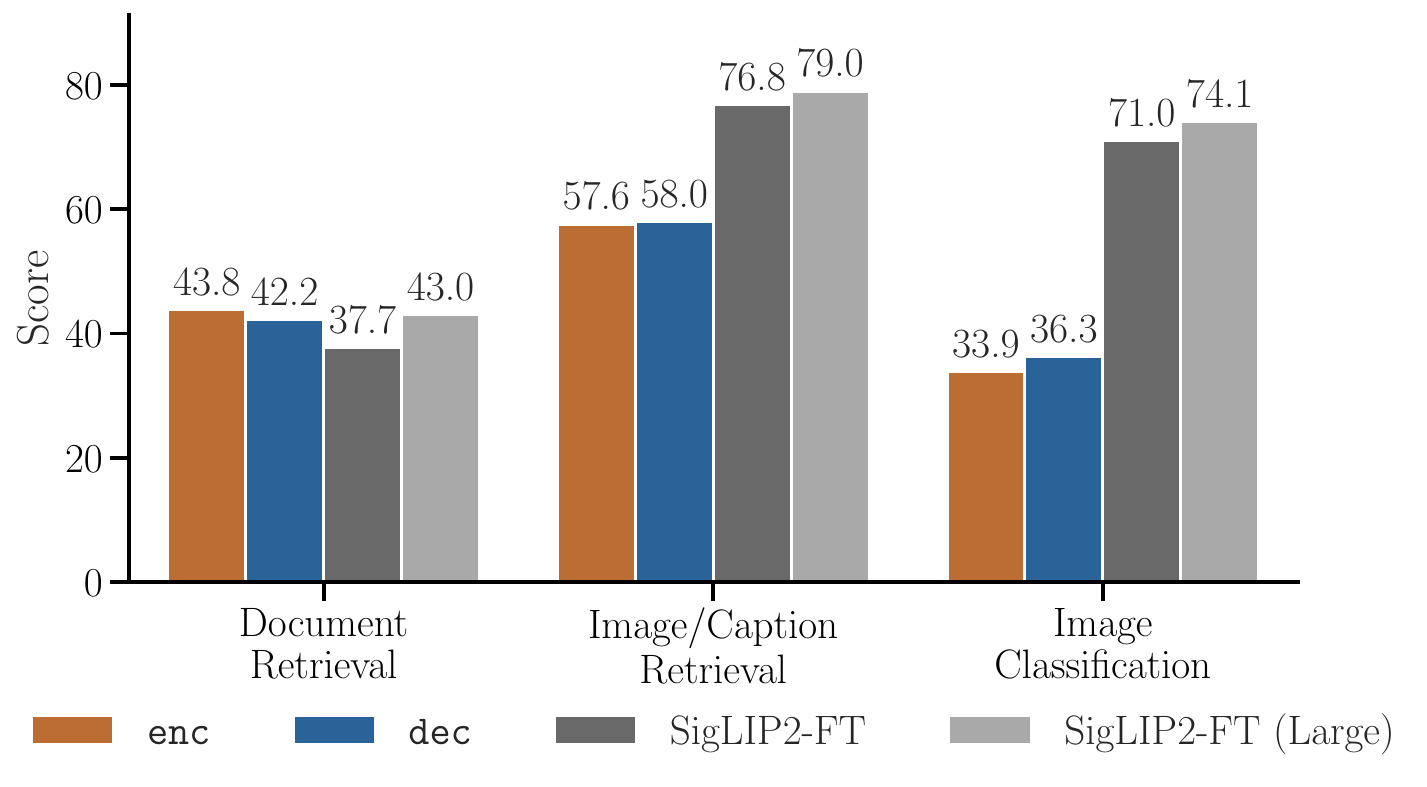}
    \caption{\textbf{Impact of Modality Alignment objective on downstream tasks.} Early Fusion of vision and text models boosts document retrieval tasks regardless of the LM objective, but degrades natural image and classification tasks w.r.t. the standalone \textit{fine-tuned} vision model SigLIP. Reported scores are aggregated MIEB scores (nDCG, Accuracy.)}
    \label{fig:objective_downstream}
  \vspace{-15pt}                         
\end{wrapfigure}

Vision-language retrievers built upon existing generative VLMs often inherit design choices and weights that may not be well suited for all embedding tasks. Here, we analyze these critical design choices hoping to derive clear insights for developing efficient visual retrievers. Importantly, although we assess design decisions at different stages of the training pipelines, evaluation are always done end-to-end on the final evaluation signal. 


\subsection{Modality Alignment Design}
\label{sec:ablation_tasks}

\noindent\textbf{Language modeling Modality Alignment improves document understanding.}
According to benchmarks such as MIEB~\citep{xiao2025miebmassiveimageembedding}, dual encoder models explicitly trained on contrastive image-text tasks outperform repurposed VLMs in natural image classification tasks. 
To assess this, we train an encoder and a decoder vision-language model using the methodology described in \autoref{sec:methodology} on a mix of natural image and document data (alignment and contrastive training). \textcolor{red}{We compare them with \textit{SigLIP2-FT}, the 378M dual vision encoder model whose vision component is used by the vision tower of both VLMs, and with the larger \textit{SigLIP2-FT Large} (881M parameters). Both SigLIP-FT models are finetuned in the same conditions as the VLMs, and initialized from pre-trained weights from scratch on billions of text-image pairs}\footnote{\textcolor{red}{We report the performance of the untrained \textit{off-the-shelf} SigLIP in Appendix~\ref{appendix:objective_downstream}}}.
As shown in Figure~\ref{fig:objective_downstream}, the two early fusion VLM variants severely underperform the SigLIP2-FT dual encoders on natural image tasks. In contrast, they achieve significant gains on document retrieval tasks (+6.1 nDCG@5 on ViDoRe and ViDoRe v2 datasets w.r.t. base), even edging out \textit{SigLIP2-FT Large} that contains 3.5x vision parameters more than both VLMs.


This confirms large-scale contrastive training remains best for high-level image representation tasks (natural images), but sequentially combining a vision model with a pretrained language model facilitates document representation tasks, even with significantly less contrastive post-training. \textcolor{red}{As the rest of this paper shows, steering away from the dual encoder architecture further enables improving performance through many avenues other than text to image contrastive training, for which supervised training samples can be hard to obtain.}

\begin{figure}[t]
    \centering    
    \includegraphics[width=0.95\textwidth]{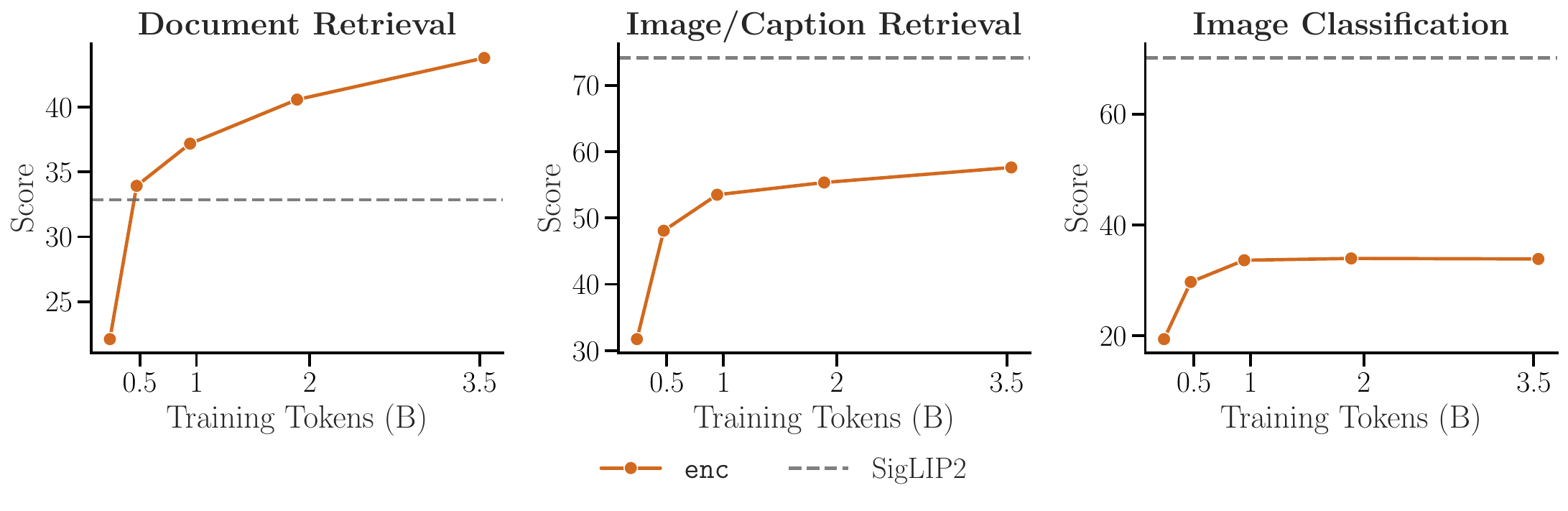}
    \caption{\textbf{Modality alignment scaling of early fusion encoders for up to 1 epoch (3.5B tokens) of data.} The dashed line indicates the vision encoder evaluated standalone without further training. Our findings show that retrieval tasks benefits from extended modality alignment phase, particularly in document retrieval, where performance quickly surpasses that of the standalone vision encoder. }
    \label{fig:pt_scaling}
\end{figure}

\noindent\textbf{Scaling the modality alignment phase for better token representations.}  
Prior work shows that scaling the modality alignment phase of VLMs improves their generative abilities~\citep{beyer2024paligemmaversatile3bvlm, mckinzie2024mm1methodsanalysis, wang2024qwen2vlenhancingvisionlanguagemodels}. We test whether similar gains hold in retrieval by contrastively fine-tuning \texttt{enc} checkpoints during MLM modality alignment. Figure~\ref{fig:pt_scaling} illustrates the results of post-trained checkpoints on diverse tasks. 
Although document retrieval improves consistently with more modality alignment data -- largely surpassing the vision tower evaluated in isolation and showing clear scaling benefits -- natural image tasks plateau past 1B tokens, far from the standalone dual encoder baseline. This shows that document and natural image retrieval leverage different mechanisms and should not be optimized the same way. \textit{Document Retrieval benefits from learning fine-grained interactions between image and text tokens through the language model, while the LM has limited utility for high level natural image tasks}.

\begin{figure*}[t]
    \centering
    \begin{minipage}{0.48\textwidth}
        \centering
        \includegraphics[width=0.9\textwidth]{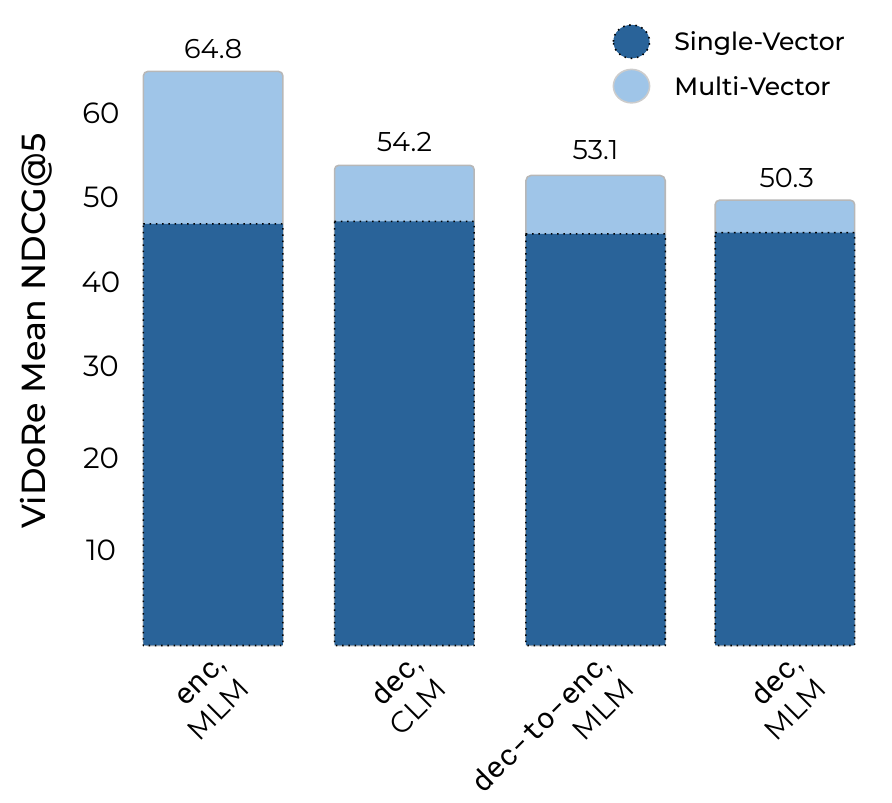}
        \caption{\textbf{Impact of attention masks and training objectives on document retrieval performances.} 
        We report the average nDCG@5 on English splits of ViDoRe benchmarks for models post-trained on ColPali.
        } 
        \label{fig:enc_dec_multivec}
    \end{minipage}%
    \hfill
    \begin{minipage}{0.48\textwidth}
        \renewcommand{\arraystretch}{1.1}
        \small
        \resizebox{\textwidth}{!}{%
        \begin{tabular}{@{} l c ccc >{\columncolor{avgcol}}c}
        & {HR Cooldown} & \rot{Document Retrieval} & \rot{Image/Caption Retrieval} & \rot{Image Classification} & \rot{\textbf{Average}} \\
    
        \midrule
        512px & \xmark & 30.7 & \underline{58.8} & \textbf{41.4} & 43.6 \\
        1024px & \xmark & 42.2 & \textbf{58.9} & \underline{37.2} & \textbf{46.1} \\
        2048px & \xmark & \underline{43.8} & 57.6 & 33.9 & 45.1 \\
        2048px & \cmark & \textbf{45.8} & 57.8 & 33.7 & \underline{45.8} \\
        
        \bottomrule
        \end{tabular}%
        }
        \captionof{table}{\textbf{Effect of image resolution on VL encoder abilities.} Document retrieval performance increases with higher image resolution. Further annealing the encoder on high-resolution images (HR Cooldown) at the end of modality alignment yields additional gains. By contrast, for non-document tasks, raising the resolution tends to degrade performance.}
        \label{table:image_resolution}
            
    \end{minipage}
\end{figure*}

\noindent\textbf{Bidirectional attention fully unlocks Late Interaction.}
Inspired by the effectiveness of bidirectional attention in text-only retrieval~\citep{gisserotboukhlef2025pretrainencodersmaskedlanguage, weller2025seqvsseqopen}\footnote{\citet{chen2025mocamodalityawarecontinualpretraining} investigate post-hoc removal of the attention mask during visual retrieval fine-tuning.}, we investigate if it surpasses causal attention in \textit{visual document retrieval}, particularly when using the multi-vector late interaction matching common in SOTA visual retrievers~\citep{khattab_colbert_2020,faysse2025colpaliefficientdocumentretrieval}.
Figure~\ref{fig:enc_dec_multivec} reports single vector and late interaction results on the ViDoRe benchmark for various model variants. On top of the standard \texttt{enc} (MLM) and \texttt{dec} (CLM) models, we evaluate the \texttt{dec-enc} and the  \texttt{dec} models modality aligned with MLM objectives to determine whether bidirectional attention capabilities can be obtained in later stages of training.

Single-vector embedding results are close between bidirectional and causal attention models for document retrieval, with \texttt{enc} slightly outperforming \texttt{dec} by +1.6 nDCG@5. 

Intuitively however, bidirectional attention makes a huge difference when used in late interaction settings, substantially exceeding the causal counterpart by +10.6 nDCG@5. 
Causal decoders are incapable of correctly contextualizing image or text token representations seen at the beginning of the sequences. \textit{This is a key result as almost all current visual retrievers, including late interaction variants, are causal models, clearly indicating some performance is left on the table.}

Removing the causal attention mask during training does not suffice to recover the \texttt{enc} late interaction performance at these data regimes. This indicates converting trained decoders as late interaction retrievers is highly non trivial, and confirms the insights from \citet{weller2025seqvsseqopen}; when possible, training encoder models from scratch remain better for retrieval tasks.

\subsection{Contrastive Training Design} 

The previous subsection established bidirectional encoder models to often be the best option when training visual retrievers. In the following experiments, we assess contrastive training choices and only report results for the encoder model for simplicity.



\noindent\textbf{Image resolution benefits are task-specific.} 
Image resolution plays a critical role in VLM generative capabilities, notably in document-focused tasks, as higher-resolution inputs enables the model to capture finer visual cues~\citep{hu2024mplugdocowl2highresolutioncompressingocrfree,marafioti2025smolvlmredefiningsmallefficient}.
Modality alignment is done at a fixed image resolution of 1024 pixels (longer side) and we report scores of contrastive training runs with varying settings in Table~\ref{table:image_resolution}. \textcolor{red}{To vary the resolution, images of the highest quality available are scaled to the desired size (often downscaled) before being fed to the image tokenizer.} Our findings confirm that embedding tasks are strongly sensitive to image-resolution. In particular, \textit{training with higher resolution inputs substantially improves the results on visual document retrieval benchmarks}, consistent with prior work in generative settings~\cite{beyer2024paligemmaversatile3bvlm,mckinzie2024mm1methodsanalysis}. Furthermore, adding a cool-down phase by showing higher-resolution images towards the end of the modality alignment phase yields additional gains. This suggests that models can adapt their attention mechanisms to finer details when exposed to increased resolution. Interestingly, these findings do not hold in natural image tasks, where high resolution can even degrade performance.

\begin{table*}[b]
  \centering
  \renewcommand{\arraystretch}{1.1}
  \small
  \resizebox{\textwidth}{!}{%
  \begin{tabular}{@{} >{}l  ccc >{\columncolor{avgcol}}c}
    & {Document Retrieval} & {Image/Caption Retrieval} & {Image Classification} & {\textbf{Average}} \\

    \midrule
    Baseline CL Mix & 43.9 & \textbf{57.2} & 36.1 & 45.7 \\
    + \textit{Text$\rightarrow$Text Pairs} & 45.6 & 53.2 & 35.7 & 44.8 \\
    + \textit{Image$\rightarrow$Caption Pairs} & \textbf{45.8} & 54.4 & \textbf{49.9} & \textbf{50.0} \\
    
    \bottomrule
  \end{tabular}%
  }
  \caption{\textbf{Impact of contrastive training mixtures on downstream tasks.} Incorporating text-only pairs improves performance on document retrieval, but degrades other performances. Adding natural images-captions pairs substantially enhances performance on classification tasks.}
  \label{table:cl_mix_impact}
\end{table*}

\noindent\textbf{Increasing the pool of contrastive pairs.}
\label{sec:scaling=-cont}
A severe limitation that current visual retrievers face is the lack of large volumes of high quality (document image, query pairs). 
Previous work \citep{ma2024unifyingmultimodalretrievaldocument, faysse2025colpaliefficientdocumentretrieval, jiang2025vlm2vectrainingvisionlanguagemodels, zhang2025gmeimprovinguniversalmultimodal} has relied on a mix of repurposed existing visual question answering datasets and synthetically generated queries with external LLMs. Even put together however, the field is only a year old, and these datasets remain small in size and often of poor quality.

A central question in our study is whether the abundance of \emph{text-only} query–document pairs can be exploited to improve \emph{visual} retrieval via cross-modal capability transfer. To probe this, we run contrastive training under three regimes. Unlike prior work that “warms up” visual retrievers or trains exclusively with text-only pairs \citep{ma2024unifyingmultimodalretrievaldocument, jiang2024e5vuniversalembeddingsmultimodal}, we \emph{interleave} text-only pairs and text–image pairs throughout training at a 1:1 ratio. The dataset sources are detailed in Appendix~\ref{appendix:contrastive_mix}

As reported in Table~\ref{table:cl_mix_impact}, incorporating text-only pairs yields a sizeable improvement on visual document retrieval (+1.7 \textsc{nDCG}@5), indicating clear cross-modal transfer—likely facilitated by the backbone’s jointly learned text–image embedding space. This result suggests that domain-specific training corpora can be assembled irrespective of native modality, reducing duplication of effort and lowering data-collection costs.

We further evaluate training with \natcap{}, a corpus of natural images paired with synthetic, highly detailed captions (see Appendix~\ref{appendix:natcap}). This scaling step improves downstream performance across the board—most notably on natural-image tasks, and with a smaller but consistent gain on document retrieval (+0.2 \textsc{nDCG}@5). Together, these findings underscore the importance of scaling contrastive learning with high-quality data, but which doesn't need to be exclusively image document focused.

%% file: sections/flagship.tex
\section{Building a Small yet Mighty Visual Retriever.}




\subsection{Training.} 

\noindent\textbf{Recipe.} Putting together the results from our experiments, we devise a training recipe for a small visual document retriever \flagship{}. It combines a state-of-the-art 150M text bidirectional encoder~\citep{weller2025seqvsseqopen} with the ModernBERT architecture \citep{modernbert} and a small vision encoder SigLIP2-16B-512 of 100M parameters~\citep{tschannen2025siglip2multilingualvisionlanguage}. We modality align both models with a MLM objective for 10B tokens, 3 times longer than during our experiments. To boost document understanding, we augment the input image resolution from 1024px to 2048px during a modality alignment cooldown stage (2B tokens). We call the resulting model \flagship{}.
Following the findings of Section~\ref{sec:scaling=-cont}, we then scale the contrastive training mix from previous experiments to combine document–query pairs with text-only pairs, and use 1 hard negatives for each document-query pair and 2 for each text-only pairs. We opt for a 2/1 text-to-image ratio following our ablation results introduced in Appendix~\ref{appendix:text_image_ratio}. This results in \colflagship{}, a compact late interaction model. For reference, we also train \textit{Bi}\flagship{}, a single vector variant. More training details are provided in Appendix~\ref{appendix:implementation_details}.

\subsection{Results.}
\label{sec:results}

\input{tables/vidore_ldb}

\noindent\textbf{\colflagship{}}. The resulting model, \colflagship{} showcases strong performances on visual document retrieval benchmarks, especially relative to its size category (Figure~\ref{fig:pareto_vidore}). Despite having over 10 times less parameters than models such as ColPali released only a year ago, it is only 0.6 nDCG@5 points below on the aggregated ViDoRe benchmark scores (\autoref{table:res}). It also edges many larger single-vector repurposed VLM models released within the year \citep{chen2025mocamodalityawarecontinualpretraining, jiang2024e5vuniversalembeddingsmultimodal, jiang2025vlm2vectrainingvisionlanguagemodels}. It however falls short of top model performance on ViDoRe which are built on larger decoder VLMs pretrained and aligned on billions of tokens of text and image data.

Most sub-1B parameter models evaluated on document retrieval benchmarks are dual encoder models, since early fusion generative models that perform well are not common at this scale. The most related model is a 176M late interaction model, ColFlor \citep{masrycolflor}, trained from the Florence2 model \citep{xiao2023florence2advancingunifiedrepresentation}. ColFlor is 12.7 nDCG@5 points under \colflagship{}. \colflagship{} also largely outperforms off-the-shelf dual encoders, even when those have substantially larger parameter counts. 
These results highlights the benefits of multi-phase training and early fusion architectures for multi-modal document related tasks, even at smaller parameter counts. We also attribute the strong performance of \colflagship{} at smaller model sizes to the symbiosis of native bidirectional attention and Late Interaction matching, which largely boosts performance relative to comparable decoder models (Section~\ref{sec:ablation_tasks}).


\noindent\textbf{Speed.} As noted by \citet{xiao2025metaembedscalingmultimodalretrieval}, multi-vector visual retrievers are not bottlenecked in their inference speed by the late interaction matching operation, but rather by the latency required to encode queries with the text model. Our model demonstrates that strong performance is not incompatible with speed, even when running inference on consumer CPUs, which is the standard setting in most industrial local deployments of text embedding models. Latencies are computed by averaging query encoding times of all NanoBEIR queries, which are 23.4 word and 147.5 character long on average, and \textcolor{red}{are run with batch size 1 to replicate online use cases. To prevent RAM bottlenecks, we benchmark on very high RAM (2TB) CPU cloud environments, but note models larger than 3B parameter require more than 12 GB RAM to run optimally.}\footnote{\textcolor{red}{With more standard CPU RAM settings such as those found in low-end servers or Google Colab (12GB RAM), models above 3B parameters must rely on memory offloading to run, which adds up to dozens of seconds of latency per query.}}  (\autoref{table:res}). \flagship{} achieves more than a 7x speedup on CPU over models with similar performances on ViDoRe. \textcolor{red}{We further report model latency results on GPU hardware in Appendix~\ref{appendix:gpu_latency}. We notably demonstrate that with batched inference, \flagship{} based query encoders are able to encode 5000 queries per second on Nvidia H100 GPUs.} ModernVBert's small model size also enables efficient batching when encoding documents.

%% file: tables/vidore_ldb.tex
\begin{table*}[t]
  \centering
  \renewcommand{\arraystretch}{1.03}
  \small
  \resizebox{\textwidth}{!}{%
  \begin{tabular}{@{} >{}l cc cc >{\columncolor{avgcol}}c >{\columncolor{latcol}}c}
    & {Late Interaction} & {Model Size (B)} & \rot{ViDoRe(v1)} & \rot{ViDoRe(v2,eng)} & \rot{\textbf{Average}} & \rot{\textbf{Latency (ms)}} \\

    \midrule
    \multicolumn{7}{@{}l}{\itshape $\ge$ 1B Parameters}\\
    MoCa-3B~\citep{chen2025mocamodalityawarecontinualpretraining}            &  & 3.75 & 80.1 & 53.8 & 66.9 & 158 \\    VLM2Vec~\citep{jiang2025vlm2vectrainingvisionlanguagemodels}             &  & 4.15 & 49.8 & 36.5 & 43.1 & 211 \\
    GME-Qwen2~\citep{zhang2025gmeimprovinguniversalmultimodal}               &  & 8.29 & 89.9 & 61.8 & 75.8 & 412\\
    E5-V~\citep{jiang2024e5vuniversalembeddingsmultimodal}                   &  & 8.36 & 62.7 & 49.4 & 56.1 & 434 \\
    ColPali~\citep{faysse2025colpaliefficientdocumentretrieval}              & \cmark & 2.92 & 81.6 & 56.8 & 69.2 & 175 \\
    ColQwen2.5~\citep{faysse2025colpaliefficientdocumentretrieval}           & \cmark & 3.75 & 89.5 & 61.5 & 75.5 & 158 \\
    Jina-v4~\citep{günther2025jinaembeddingsv4universalembeddingsmultimodal} & \cmark & 3.75 & 90.4 & 60.1 & 75.2 & 158 \\
    NemoRetriever-3B~\citep{xu2025llamanemoretrievercolembedtopperforming}   & \cmark & 4.40 & 91.0 & 66.3 & 78.7 & 155 \\

    \midrule
    \multicolumn{7}{@{}l}{\itshape $\le$ 1B Parameters}\\
    Jina CLIP$^*$~\citep{koukounas_jina_2024}                                    &  & 0.22 & 17.6 & 14.0 & 15.8 & \textbf{14} \\
    BGE Visualized M3$^*$~\citep{zhou2024vistavisualizedtextembedding}           &  & 0.87 & 12.4 & 10.2 & 11.3 & 38 \\
    SigLIP2-L-512/16$^*$~\citep{tschannen2025siglip2multilingualvisionlanguage}  &  & 0.88 & 43.8 & 27.0 & 35.4 & 25 \\
    ColFlor~\citep{masrycolflor}                                             & \cmark & 0.17 & 68.8 & 43.0 & 55.9 & 17 \\
    \biflagship{} (ours)                                           &  & 0.25 & 63.6 & 35.7 & 49.7 & 20 \\
    \textbf{\colflagship{} (ours)}                                           & \cmark & 0.25 & \textbf{81.2} & \textbf{56.0} & \textbf{68.6} & 20 \\
    
    \bottomrule
  \end{tabular}%
  }
  \caption{\textbf{Performance on ViDoRe.}
  Our model \colflagship{} offers the best performance-size tradeoff, significantly outperforming existing sub-1B models and matching the performance of models up to 10x larger with substantially lower inference CPU latency \textcolor{red}{Details and GPU latencies in Appendix \ref{appendix:gpu_latency}. Models marked with $^*$ are not specifically trained for VDR.} Bold values indicate the best performance amongst sub-1B models.}
  \label{table:res}
\end{table*}

%% file: sections/related_work.tex
\section{Related Work}

\noindent\textbf{Repurposing VLMs for Representation Learning.}  
Motivated by the zero-shot performances of generative VLMs~\citep{alayrac2022flamingovisuallanguagemodel, lucas_beyer_paligemma_2024, bai_qwen-vl_2023}, recent studies have explored repurposing these for multimodal embedding tasks~\citep{ma2024unifyingmultimodalretrievaldocument,faysse2025colpaliefficientdocumentretrieval,jiang2025vlm2vectrainingvisionlanguagemodels, zhang2025gmeimprovinguniversalmultimodal}. As backbone generative models improved, retriever performance improved as well showcasing the central impact of language model pretraining and modality alignment \citep{xu2025llamanemoretrievercolembedtopperforming, nussbaum2025nomicembedtrainingreproducible}. 
These model remain inherently constrained by their causal attention mechanisms which has been shown in text settings to limits represational efficiency ~\citep{gisserotboukhlef2025pretrainencodersmaskedlanguage,weller2025seqvsseqopen}. Recent work attempts to address this issue by modifying VLM attention during continual pretraining \citep{chen2025mocamodalityawarecontinualpretraining} or contrastive tuning ~\citep{jiang2025vlm2vectrainingvisionlanguagemodels, xu2025llamanemoretrievercolembedtopperforming}, but no recent work attempts to align natively bidirectional language encoder models with vision encoders. The recent release of long sequence text encoders \citep{modernbert, boizard2025eurobertscalingmultilingualencoders} makes this possible.

\noindent\textbf{Late Interaction in Visual Document Retrieval}  To further boost performance, visual document retrievers leverage the late interaction mechanism~\citep{khattab_colbert_2020} which matches multiple query embeddings with multiple document embeddings through the MaxSim operation ~\citep{faysse2025colpaliefficientdocumentretrieval,günther2025jinaembeddingsv4universalembeddingsmultimodal, xu2025llamanemoretrievercolembedtopperforming}. This enables more granular interactions between image and query tokens, at the cost of additional storage and a slight compute overhead during the matching operation. Efficiency gains have come from improving the storage costs through quantization \citep{vespaScalingColPali}, token pruning \citep{faysse_croissantllm_2024} and more recently the use of Matrioshka losses to compact multi-token representations \citep{xiao2025metaembedscalingmultimodalretrieval}. Ultimately, the performance bottleneck when running visual retrieval inference with such models now resides mostly in the necessity to rely on costly GPU hardware to encode queries, which sets apart text from vision retrieval. This paper fills this gap by using encoders that run on CPU, of parameter sizes comparable to commonly used local text embedding models \citep{chen_bge_2024, enevoldsen2025mmtebmassivemultilingualtext}.





%% file: sections/conclusion.tex
\section{Conclusion}

In this paper we question design decisions of current VLM-based retriever models, providing crucial insights into what matters when training early-fusion vision encoders. Our study notably shows that these models generally do not improve retrieval on natural-image tasks compared to dual encoders, yet strong vision-language alignment is essential for document-centric retrieval. We uncover a tight synergy between bidirectional attention and late-interaction retrieval, which underscores a fundamental limitation of repurposing decoder-style generative VLMs for retrieval. To mitigate data scarcity in contrastive learning, we propose augmenting limited image-document/text-query pairs with larger, lower-cost corpora from other modalities. Guided by these insights, we trained \colflagship{}, a compact yet powerful $250$M-parameter multimodal encoder that matches the performance of models up to $10\times$ larger on visual retrieval benchmarks. We release models and training code to help practitioners reduce cost and latency when deploying visual retrievers in real-world applications, and to encourage research on efficient multimodal embedding models.

\noindent\textbf{Future Work \& Limitations.}  
By design, our analysis targets relatively small models. An important next step is to test whether the observed patterns persist at larger scales—for example, to more rigorously probe the interplay between late interaction and bidirectional attention. Our study also focuses exclusively on English. While we expect the broad trends to generalize and see clear value in releasing multilingual variants, it remains unclear how allocating parameters to additional languages trades off against the understanding of the vision modality, and to what extent this penalizes English retrieval performance as the number of languages are scaled \citep{pmlr-v202-fernandes23a}. Finally, although we center on retrieval and sequence-level zero-shot classification, the modality-aligned encoder can be fine-tuned for a range of token-level tasks, including OCR error detection, token-level classification, visual named entity recognition, visually grounded token-level object detection, contextual embeddings \citep{conti2025context}. We release our base model to encourage exploration of these directions.


%% file: sections/appendix.tex
\section{Training}

\subsection{Implementation and Resources}
\label{appendix:implementation_details}
\input{tables/implementation_details}
We list hyperparameters and resource details in \autoref{table:app_training_details} for the various training stages of our final models. We employ ZeRO stage 1 optimizer~\citep{rajbhandari2020zeromemoryoptimizationstraining} for our modality alignment runs. All ablation models are contrastively trained with gradient checkpointing~\citep{chen2016trainingdeepnetssublinear} to reduce memory usage. All training runs are performed with FlashAttention 2.0~\citep{dao2023flashattention2fasterattentionbetter}. For LoRA configurations, we consistently use a rank \texttt{r} of $32$, \texttt{lora\_alpha} of $32$, and a dropout of $0.1$. For the implementation, we start from m4\footnote{SmolVLM trainer, \url{https://github.com/huggingface/smollm}} and ColPali\footnote{\url{https://github.com/illuin-tech/colpali}} codebases for training, and use the MTEB\footnote{\url{https://github.com/embeddings-benchmark/mteb}} repository for evaluation.\footnote{\ghrepo}

\subsection{Similarity Functions}
\label{appendix:similarity_functions}
\noindent\textbf{Single-Vector Similarity.}
For single-vector models, we apply mean pooling for MLM-aligned encoders and end-of-sequence (EOS) pooling for CLM-based models and compute the cosine similarity of a query $q$ and a document $d$ as
\begin{equation}
\label{eq:cosine_sim}
    \Phi_\mathrm{CosSim}(\mathbf{q},\mathbf{d}) = \exp(\cos(\mathbf{E}_q,\mathbf{E}_d)/\tau)
\end{equation}
\noindent\textbf{Multi-Vector Similarity.}
For multi-vector models, we adopt the standard late-interaction scoring function defined as:
\begin{equation}
\label{eq:li_sim}
\Phi_\mathrm{LI}(q,d) = \sum_{i \in \llbracket 1, N_q \rrbracket}
\max_{j \in \llbracket 1, N_d \rrbracket}
\left\langle \mathbf{E}_q^{(i)} , \mathbf{E}_d^{(j)} \right\rangle ,
\end{equation}
where $\mathbf{E}_q^{(i)}$ and $\mathbf{E}_d^{(j)}$ denote token-level embeddings for the query and document, respectively.

\subsection{Data}

\subsubsection{Modality Alignment Mixture}
\label{appendix:alignment_mixture}

For our modality alignment trainings, we rely on The Cauldron dataset \citep{laurencon_what_2024} and its Docmatix extension \citep{laurençon2024building}. \autoref{tab:mod_alignment_mix} provides further details on the constitution of this dataset.
\input{tables/alignment_mix}

\subsubsection{\natcap{}}
\label{appendix:natcap}
\begin{figure}[t]
    \centering    
    \includegraphics[width=0.95\textwidth]{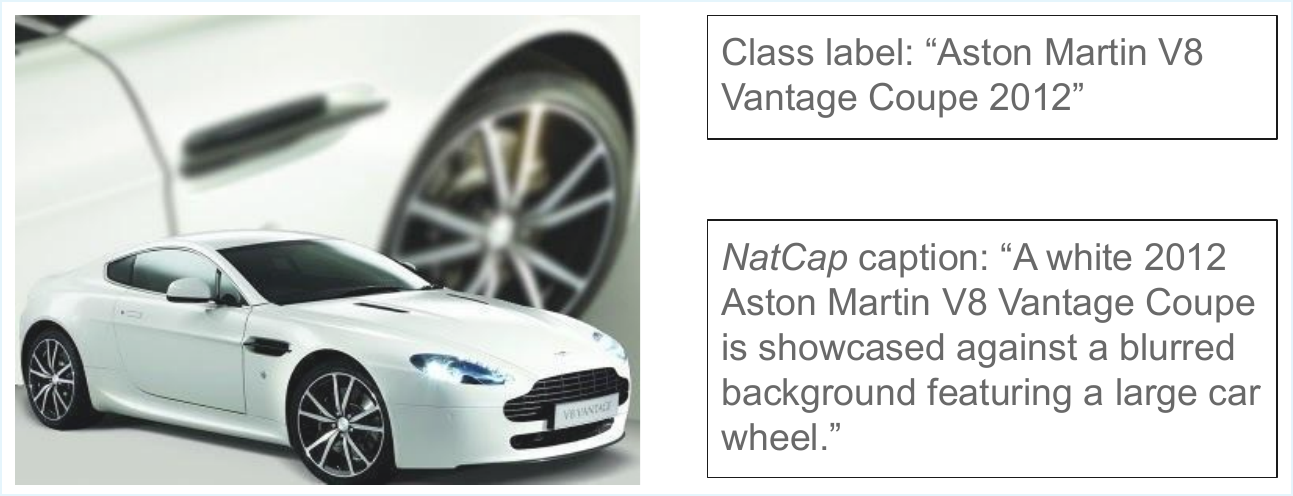}
    \caption{Example from the NatCap dataset}
    \label{fig:natcap_example}
\end{figure}

To enrich our contrastive learning data mixture, we construct \natcap{} (Natural Captions), a large-scale dataset containing around 333000 contextualized image–caption pairs. This dataset is created by generating synthetic captions, along with cross-class and in-class discriminative tags, from existing image classification datasets (see Table~\ref{tab:natcap}). For this purpose, we leverage \texttt{Gemini-flash-2.5}\footnote{\url{https://ai.google.dev/gemini-api/docs/models?hl=fr\#gemini-2.5-flash}} which produces captions conditioned on both the image content and the accompanying dataset metadata, as illustrated in Figure~\ref{fig:natcap_example}. We detail the prompt below.

\input{tables/natcap}

\subsubsection{Contrastive Training Mix}
\label{appendix:contrastive_mix}

In this subsection, we describe the composition of our data mixes used in the contrastive training stages. \autoref{tab:cl_mix} outlines the datasets included in each mix, including the Document-Focused variant employed for \colflagship{}.

\input{tables/cl_mix}

\section{Baselines Details}
\label{appendix:baselines}
In this section, we describe the models evaluated in as comparison to our document retriever model.

\noindent{\textbf{MoCa-3B}~\citep{chen2025mocamodalityawarecontinualpretraining}.} 
A modality-aware continual pretraining model that transforms a causal vision-language model into a bidirectional multimodal embedding model, using interleaved image-text reconstruction and contrastive alignment to support cross-modal retrieval.

\noindent{\textbf{GME-Qwen2}~\citep{zhang2025gmeimprovinguniversalmultimodal}.} 
A unified multimodal embedder built on Qwen2-VL~\citep{wang2024qwen2vlenhancingvisionlanguagemodels}, which produces shared embedding representations across text, image, and fused input modalities, enabling universal multimodal retrieval.

\noindent{\textbf{VLM2Vec}~\citep{jiang2025vlm2vectrainingvisionlanguagemodels}.} 
A method that trains a vision-language encoder by converting a VLM through extensive contrastive post-training. Flagship model is based on the model Phi-3.5~\citep{abdin_phi-3_2024}.

\noindent{\textbf{E5-V}~\citep{jiang2024e5vuniversalembeddingsmultimodal}.} 
An adaptation of the E5 embedding approach to multimodal models: it trains only on text pairs yet bridges the modality gap to handle image inputs, reducing cost while achieving universal embeddings.

\noindent{\textbf{ColPali}~\citep{faysse2025colpaliefficientdocumentretrieval}.} 
A vision-based document retrieval model that processes document pages as images (no OCR) and produces multi-vector embeddings via a late-interaction mechanism over PaliGemma~\citep{beyer2024paligemmaversatile3bvlm}, enabling efficient and accurate retrieval.

\noindent{\textbf{ColQwen2.5}~\citep{faysse2025colpaliefficientdocumentretrieval}.} 
An extension of ColPali~\citep{faysse2025colpaliefficientdocumentretrieval} using Qwen2-VL~\citep{wang2024qwen2vlenhancingvisionlanguagemodels} as the backbone, carrying forward the late interaction retrieval paradigm over page image embeddings, capturing layout and textual context without OCR.

\noindent{\textbf{Jina-v4}~\citep{günther2025jinaembeddingsv4universalembeddingsmultimodal}.} 
A multimodal embedding model combining visual and textual inputs with support for multi-vector (late interaction) embeddings, using adapters over a unified backbone to excel on visually rich document retrieval.

\noindent{\textbf{NemoRetriever}~\citep{xu2025llamanemoretrievercolembedtopperforming}.} 
An LI retriever that combines vision-language embeddings with a ColEmbed design, enabling high performance on visual document retrieval with structured patch matching and efficient similarity.

\noindent{\textbf{Jina CLIP}~\citep{koukounas_jina_2024}.} 
A smaller scale vision-language model using CLIP embeddings, applied to document retrieval tasks; although not LI, it offers a lightweight multimodal baseline.

\noindent{\textbf{BGE Visualized M3}~\citep{zhou2024vistavisualizedtextembedding}.} 
A vision-enhanced version of BGE M3~\citep{chen_bge_2024} that supports visual inputs and extends embedding models into multimodal domains.

\noindent{\textbf{SigLIP2-L-512/16}~\citep{tschannen2025siglip2multilingualvisionlanguage}.} 
A multilingual vision-language bi-encoder model, which combines image and text modalities to yield unified embeddings across languages. This configuration handles images of 512x512 pixels and create subpatches of 16x16 pixels.

\noindent{\textbf{ColFlor}~\citep{masrycolflor}.} 
A lightweight OCR-free visual document retriever with only 174M parameters built over Florence-2 and DaViT, delivering strong performance near ColPali with much lower computational cost and much faster encoding.




\section{Additional Ablations}

\subsection{Performance Against Off-the-Shelf Dual Encoder}
\label{appendix:objective_downstream}
\begin{figure}
    \centering
    \includegraphics[width=0.75\linewidth]{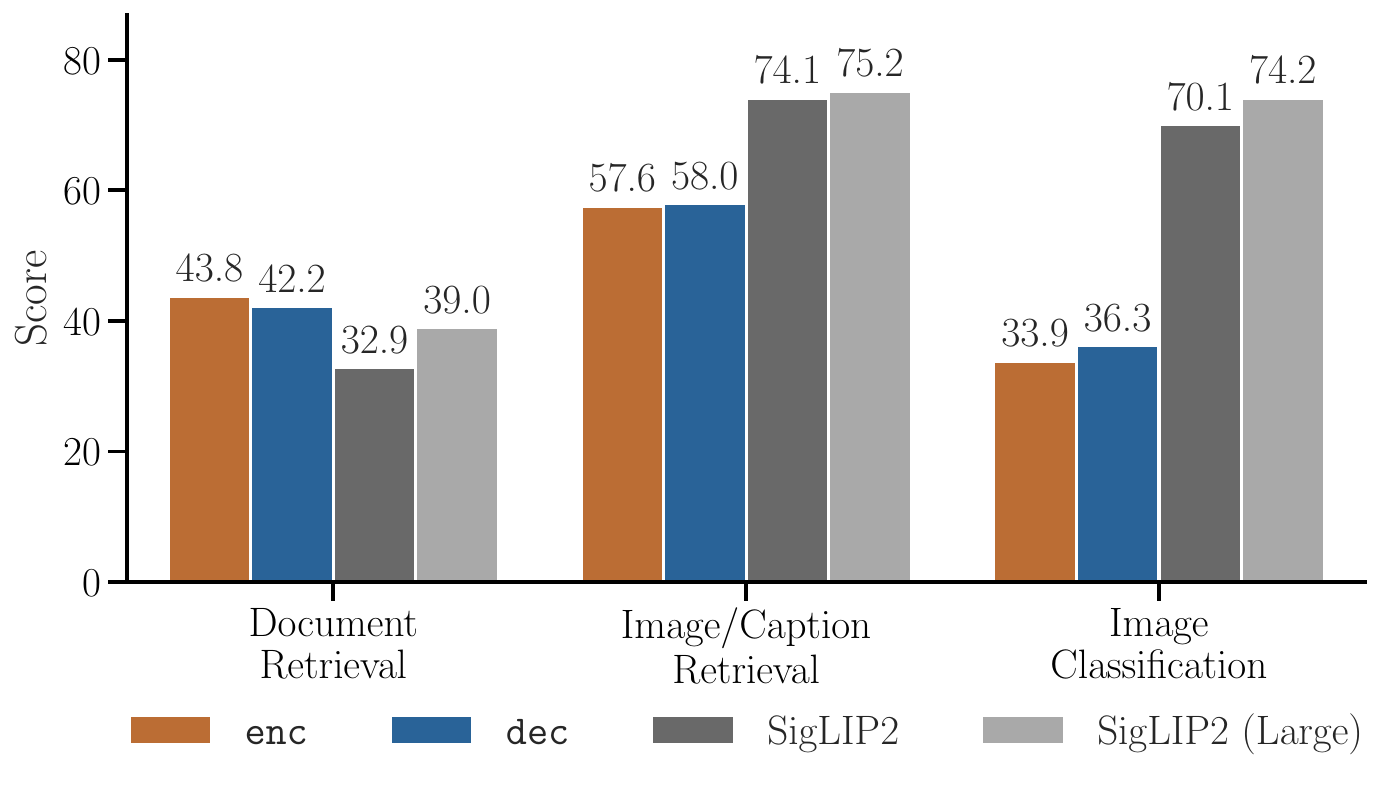}
    \caption{\textbf{Impact of Modality Alignment objective on downstream tasks.} Early Fusion of vision and text models boosts document retrieval tasks regardless of the LM objective, but degrades natural image and classification tasks w.r.t. the standalone \textit{off-the-shelf} vision model SigLIP. Reported scores are aggregated MIEB scores (nDCG, Accuracy.)}
    \label{fig:objective_downstream_noft}
\end{figure}

\textcolor{red}{We study whether using \textit{off-the-shelf} performances of the standalone vision tower are not outweighing the burden of adding language parameters and re-training through language modeling, as proposed in our work. Figure~\ref{fig:objective_downstream_noft} shows the results of the various models on the tasks described in Section~\ref{sec:methodology}. Similarly to Section~\ref{sec:ablation_tasks}, we observe that the early fusion model trained with LM objective significantly outperform the standalone vision tower on document retrieval tasks (+10.9 nDCG@5). It even surpass the larger dual encoder (+4.8 nDCG@5) on these latest tasks. We note that the standalone vision tower largely outperform the early fusion models on the other natural images tasks, supporting for the use of the SigLIP model for these tasks as found in various general benchmarks~\citep{xiao2025miebmassiveimageembedding}.}

\subsection{Scaling Dynamics of Attention Masks}
\label{appendix:mlm_vs_clm_scaling}
\begin{figure}[ht]
    \centering
    \includegraphics[width=\linewidth]{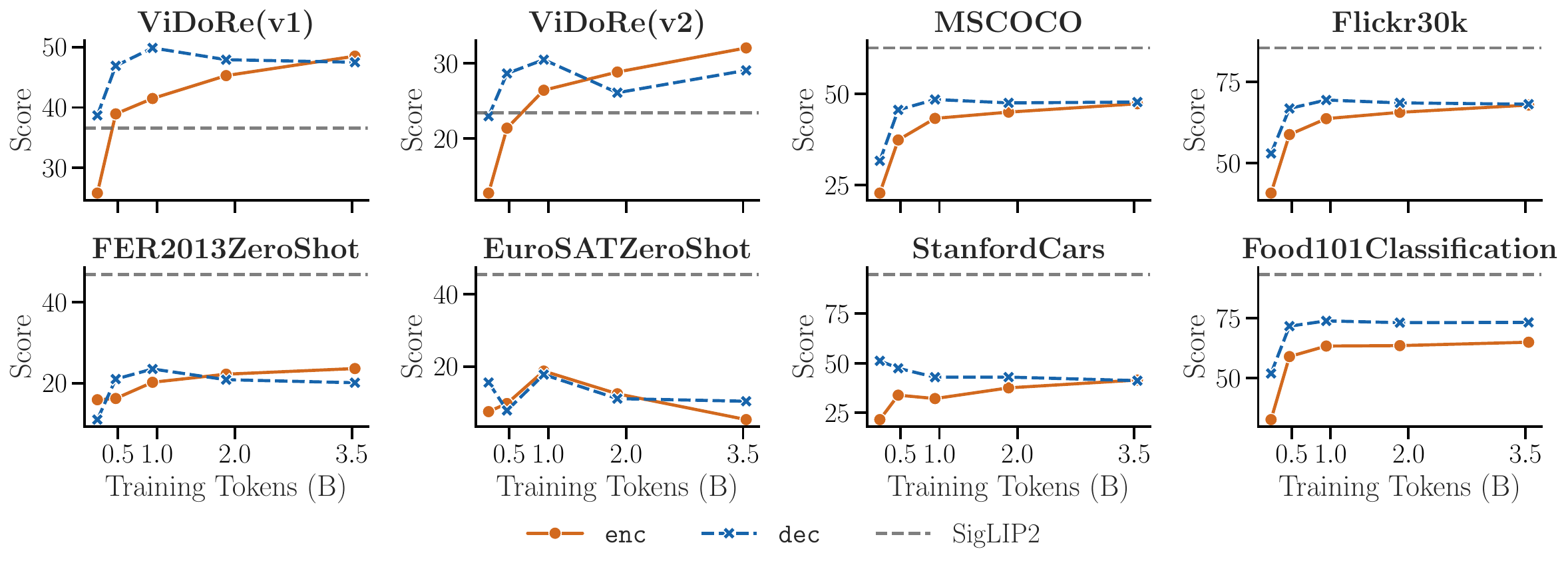}
    \caption{\textbf{Attention masks impact on modality alignment phase scaling.} The dashed line marks the vision tower baseline. The orange curve shows the model initialized from a decoder LM with a \textit{CLM} objective, and the blue curve shows the model trained with an \textit{MLM} objective from an encoder LM. CLM performs better in low-data regimes, but MLM scales more effectively, surpassing CLM in document retrieval, while captioning and classification remain below the CLIP baseline.}
    \label{fig:pt_scaling_clmvmlm}
\end{figure}

We study the different training dynamics of the different training objectives. We compare the \texttt{enc} (MLM) approach with a traditional \texttt{dec} (CLM) objective. Figure~\ref{fig:pt_scaling_clmvmlm} presents the performance of the two training objectives across a diverse set of tasks. While starting \texttt{dec} offers an advantage in low-data regimes, \texttt{enc} seems to catches up. In document retrieval tasks, it eventually surpasses \texttt{dec} and scales better.

\subsection{Bridging the Gap with Longer Contrastive Training}
\label{appendix:cl_scaling}
\begin{figure}[ht]
    \centering
    \includegraphics[width=\linewidth]{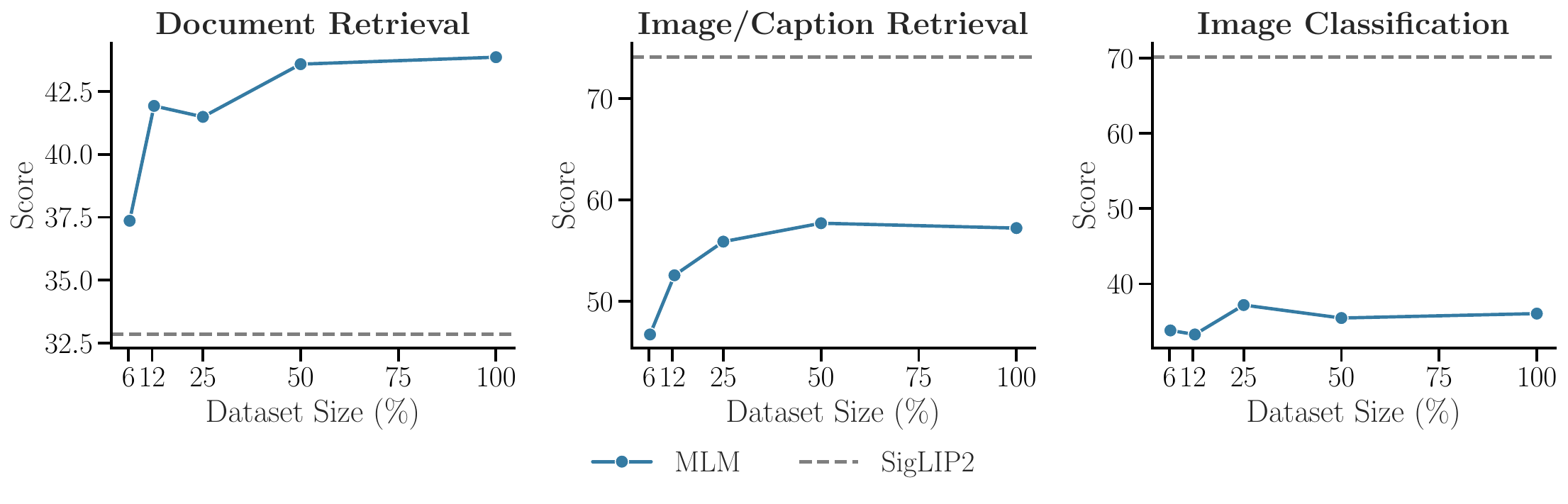}
    \caption{Contrastive training scaling. Each dot on the blue curve represents one fraction of the baseline contrastive training mix (ColPali + MSCOCO). Performance improves with more in-distribution data, surpassing the baseline on document benchmarks and narrowing the gap on image captioning. There is no clear improvement in image classification, highlighting the need for more diverse pairs.}
    \label{fig:cl_scaling}
\end{figure}

We study the impact of additional in-distribution training pairs on embedding tasks by scaling the contrastive training stage. Starting from the final checkpoint of our encoder-based ablation model, we double the contrastive dataset size at each step and train until convergence\footnote{To avoid overfitting, we set an early stopping on an eval set. We limit the number of step to one epoch on the full dataset.}. This setup tests whether scaling continues to improve performance. 
Figure~\ref{fig:cl_scaling} shows the scaling behavior. Performance improves overall with more in-distribution data. The vision-tower baseline is quickly surpassed on visual document benchmarks, and scaling narrows the gap on other tasks\footnote{Note that the models probably won't fully recover baseline vision-tower performance. This highlights the need to choose models according to use case (e.g., lightweight CLIP-like models for image classification).}. We note a plateau in captioning and classification, pointing to the need for more diverse data.  

\subsubsection{Optimal Text-To-Image Ratio for Document Retrieval}
\label{appendix:text_image_ratio}
\begin{figure}[ht]
    \centering
    \includegraphics[width=.95\linewidth]{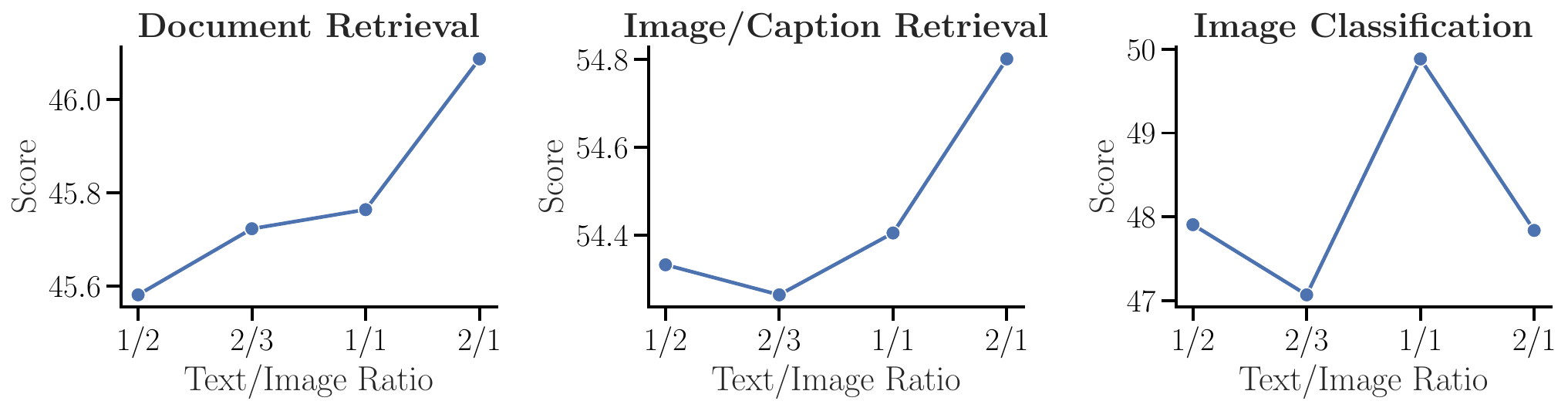}
    \caption{\textbf{Optimal text-to-image ratio in contrastive training mix.} Increasing the ratio in retrieval tasks consistently improves the performances. }
    \label{fig:text_image_ratio}
\end{figure}

Our findings in subsection~\ref{sec:scaling=-cont} indicate that incorporating additional text-only pairs boosts document retrieval performance. While our initial experiment employed a 1:1 text-to-image ratio, we further investigate how varying this ratio impacts our broad set of tasks. We start from the best contrastive mix in Table~\ref{table:cl_mix_impact}, and vary the text-to-image ratio. As shown in Figure~\ref{fig:text_image_ratio}, increasing the number of text-only pairs \textit{for a fixed amount of image pairs} consistently enhances retrieval performance. However, for natural image classification tasks, adding more text does not appear to provide benefits.

\subsection{Late Interaction for Non-Documental Retrieval}

\begin{table*}[ht]
  \centering
  \renewcommand{\arraystretch}{1.2}
  \small
  \resizebox{\textwidth}{!}{%
  \begin{tabular}{@{} >{}l c cccc >{\columncolor{avgcol}}c}
    & & \multicolumn{2}{>{\columncolor{grpA}}c}{\textbf{Document Retrieval}} & \multicolumn{2}{>{\columncolor{grpB}}c}{\textbf{Image/Caption Retrieval}} \\
    \cmidrule(lr){3-6}
& \rot{Model Size} & \rot{ViDoRe(v1)} & \rot{ViDoRe(v2)} & \rot{MSCOCO (T$\rightarrow$I)} & \rot{Flickr30k (T$\rightarrow$I)} & \rot{\textbf{Average}} \\

    \midrule
    \multicolumn{7}{@{}l}{\itshape CLIP Encoders}\\
    siglip2-base-patch16-512 & 376M & 36.6 & 23.4 & 66.2 & 86.9 & 53.3 \\
    siglip2-large-patch16-512 & 882M & 43.8 & 27.0 & 67.1 & 88.9 & 56.7 \\
    clip-vit-base-patch16 & 151M & 25.5 & 20.4 & 50.3 & 76.8 & 43.3 \\
    clip-vit-large-patch14 & 428M & 38.0 & 28.6 & 52.7 & 79.3 & 49.6 \\

    \midrule
    \multicolumn{7}{@{}l}{\itshape VLM-based Encoders}\\
    VLM2Vec-Full & 4150M & 49.8 & 36.5 & 59.5 & 81.8 & 56.9 \\
    e5-v & 8360M & 62.7 & 49.4 & 68.1 & 89.8 & 67.5 \\

    \midrule
    \multicolumn{7}{@{}l}{\itshape Early Fusion Encoders}\\
    bge-visualized-base & 196M & 10.3 & 9.0 & 50.0 & 74.1 & 35.9 \\
    bge-visualized-m3 & 873M & 12.4 & 10.2 & 39.6 & 69.0 & 32.8 \\
    \textbf{\flagshipembed{}} & 252M & 58.4 & 36.9 & 56.5 & 76.0 & 56.9 \\
    \textbf{\flagshipembed{} (multi-vector)} & 252M & 76.5 & 53.9 & 61.8 & 81.4 & \textbf{68.4} \\
    
    \bottomrule
  \end{tabular}%
  }
  \caption{\textbf{Generalist retrieval performances.} Late interaction benefits extend to non-documental retrieval tasks. Our multi-vector model increases its single-vector counterpart across all tasks, surpassing larger VLM-based retrievers. }
  \label{tab:results_generalist}
\end{table*}

We want to study if the multi-vector gains transfer to non-documental retrieval. To do so, we contrastively post-train our base model on our generalist post-training mix presented in Table~\ref{tab:cl_mix}.
The late interaction generalist exhibits superior performance in retrieval setting, improving its single-vector performance by +20.2\% (11.5 points), matching the performance of substantially larger VLM-based retrievers like E5-V (8.3B parameters, 67.5 points) and surpassing dual encoders like SigLIP (882M parameters, 56.7 points). This matches the capabilities observed in Section~\ref{sec:ablation_tasks} for documental settings for models with native bidirectional attention, extending it to natural image tasks. This result extends the prevailing understanding from the document retrieval community, where the superiority of late-interaction is well-documented (\cite{khattab_colbert_2020}, \cite{GTE-ModernColBERT}, \cite{faysse2025colpaliefficientdocumentretrieval}). While this performance gap is widely accepted for document retrieval, its applicability to caption matching tasks has not really been addressed. Our findings provide strong evidence that the fine-grained matching capabilities of late-interaction models are a key driver of performance in this domain too.

\subsubsection{Model Merging}\label{appendix:model_merging}
\begin{figure}[ht]
    \centering
    \includegraphics[width=.7\linewidth]{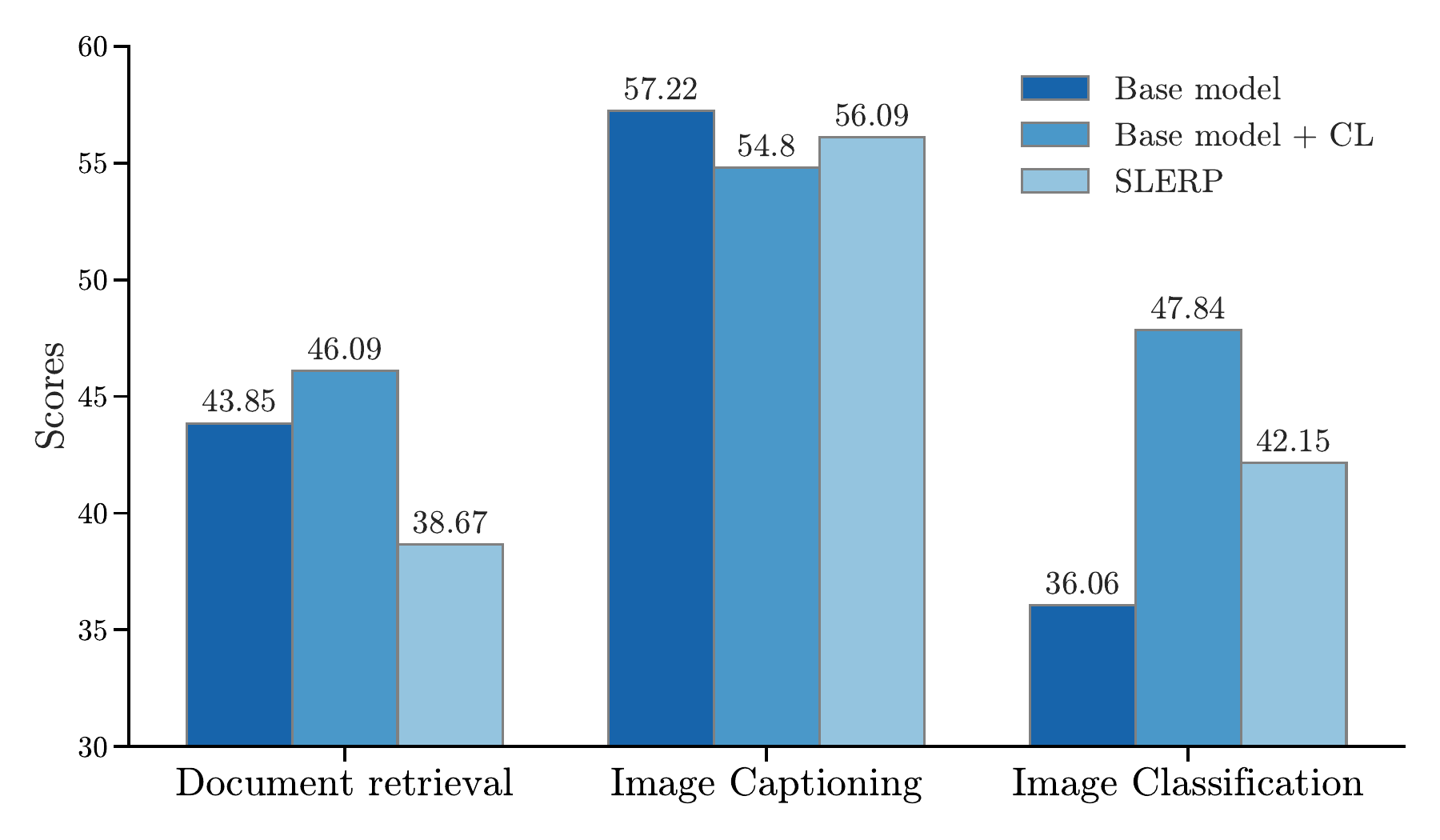}
    \caption{Merging model results across tasks. Benefits are task-dependent, with performance degradation w.r.t. both original models in Document Retrieval.}
    \label{fig:model_merging_results}
\end{figure}

Our contrastive learning stage provides direct performance trade-offs on different tasks. Following recent trends, we evaluate how model merging techniques allow to mitigate performance degradation on specific tasks, while maintaining the performance enabled by the contrastive training \citep{sung2023empiricalstudymultimodalmodelmerging, dziadzio2024mergemultimodalmodelstime, li2024improvinggeneraltextembeddingmerging, zhang2025qwen3embeddingadvancingtext}. We merge our ablation model after modality alignment with the checkpoint after the full contrastive learning with two methods: SLERP \citep{ilharco2022patchingopenvocabularymodelsinterpolating} and average merging \citep{slerp_original}. For SLERP, we compare three values for the $\lambda$ coefficient ($0.25$, $0.5$, $0.75$). \autoref{fig:model_merging_results} displays the the trends with the best method (SLERP, $\lambda = 0.75$). As we can see, the merged model mitigates the performance drop in Image/Caption Retrieval tasks, while maintaining significant gains on Image Classification tasks. However, merging strongly degrades performance on Document Retrieval, showing that benefits of merging embedding models are task-dependent. 

\subsubsection{Curriculum For Document Retriever Contrastive Post-Training}

\begin{wraptable}{r}{0.5\textwidth}
\vspace{-10pt}
\centering
\renewcommand{\arraystretch}{1.2}
\small
\resizebox{\linewidth}{!}{%
  \begin{tabular}{@{} l cc >{\columncolor{avgcol}}c}
    & \rot{ViDoRe(v1)} & \rot{ViDoRe(v2)} & \rot{\textbf{Average}} \\
    \toprule
    \multicolumn{4}{@{}l}{\itshape Document retrieval contrastive training starting checkpoint}\\
    \midrule
    \flagship{}-base                        & \textbf{81.2} & \textbf{56.0} & \textbf{68.6} \\
    \hspace{1cm} + multi-vector generalist CL  & 80.7 & 55.4 & 68.1 \\
    \hspace{1cm} + single-vector generalist CL   & 80.6 & 54.0 & 67.3 \\
    \bottomrule
  \end{tabular}
}
\caption{\textbf{Performance of \flagship{} Document Specialisation Curriculums.} This table presents the performance of various contrastive training curriculums starting from \flagship{}-base, on the ViDoRe(v1) and ViDoRe(v2) benchmarks. The generalist contrastive learning mix used in the last two models is detailed in Table~\ref{tab:cl_mix}. We see that a preliminary stage of generalist contrastive learning harms the final document retrieval performance, regardless of whether a multi-vector approach is used.}
\label{appendix:flagship_ablations}
\end{wraptable}

We conduct an ablation study to determine the optimal contrastive training curriculum for specializing \flagship{} in document retrieval. Specifically, we investigate whether a preliminary generalist contrastive training phase, intended to leverage a larger dataset, improves downstream performance. As shown in Table \ref{appendix:flagship_ablations}, our results demonstrate that this initial generalist phase is detrimental to final performance ($-0.5\%$). The optimal strategy is to specialize the model on the target task directly after its initial Masked Language Modeling (MLM) alignment.

\subsection{Text-Only Retrieval}
\begin{table}[h]
\centering
\begin{tabular}{l r r}
  \toprule
  Model & Params (M) & NDCG@5 \\
  \midrule
  \multicolumn{3}{l}{\textbf{Statistical}} \\
  BM25s & — & 0.559 \\
  \midrule
  \multicolumn{3}{l}{\textbf{Single Vector}} \\
  Jina Embeddings v4 & 3577* & 0.623 \\
  E5-large-v2 & 335 & 0.605 \\
  bge-m3 (Bi Encoder) & 567 & 0.590 \\
  Qwen3-Embedding-0.6B & 600 & 0.567 \\
  \midrule
  \multicolumn{3}{l}{\textbf{Multi Vector}} \\
  LightOn GTE-ModernColBERT v1 & 149 & 0.669 \\
  Jina ColBERT v2 & 137 & 0.642 \\
  bge-m3 (Late Interaction) & 567 & 0.606 \\
  ColBERT v2 & 110 & 0.593 \\
  Colqwen2-v1.0 & 1580* & 0.593\\
  \colflagship{} & 150* & 0.589 \\
  Colqwen2.5-v0.2 & 3145* & 0.589\\
  \bottomrule
\end{tabular}
\caption{Average NDCG@5 of \colflagship{} on NanoBEIR, a text retrieval benchmark with multiple sub domains. *For multimodal models, we only consider parameters of the text encoder}
\label{appendix:text_perf}
\end{table}

The results in Table \ref{appendix:text_perf} detail the performance of \colflagship{} and other baselines on the NanoBEIR text retrieval benchmark. It achieves an average NDCG@5 score competitive with single and multi vector models specialized for text, even without explicit optimization for this modality. This performance is encouraging and indicates a promising direction for training a unified model for both text and image retrieval.

\subsection{Model Latency}

\subsubsection{Image Resolution Tradeoffs}
\label{appendix:visual_tokens}

\begin{figure}[ht] 
\centering \includegraphics[width=0.6\linewidth]{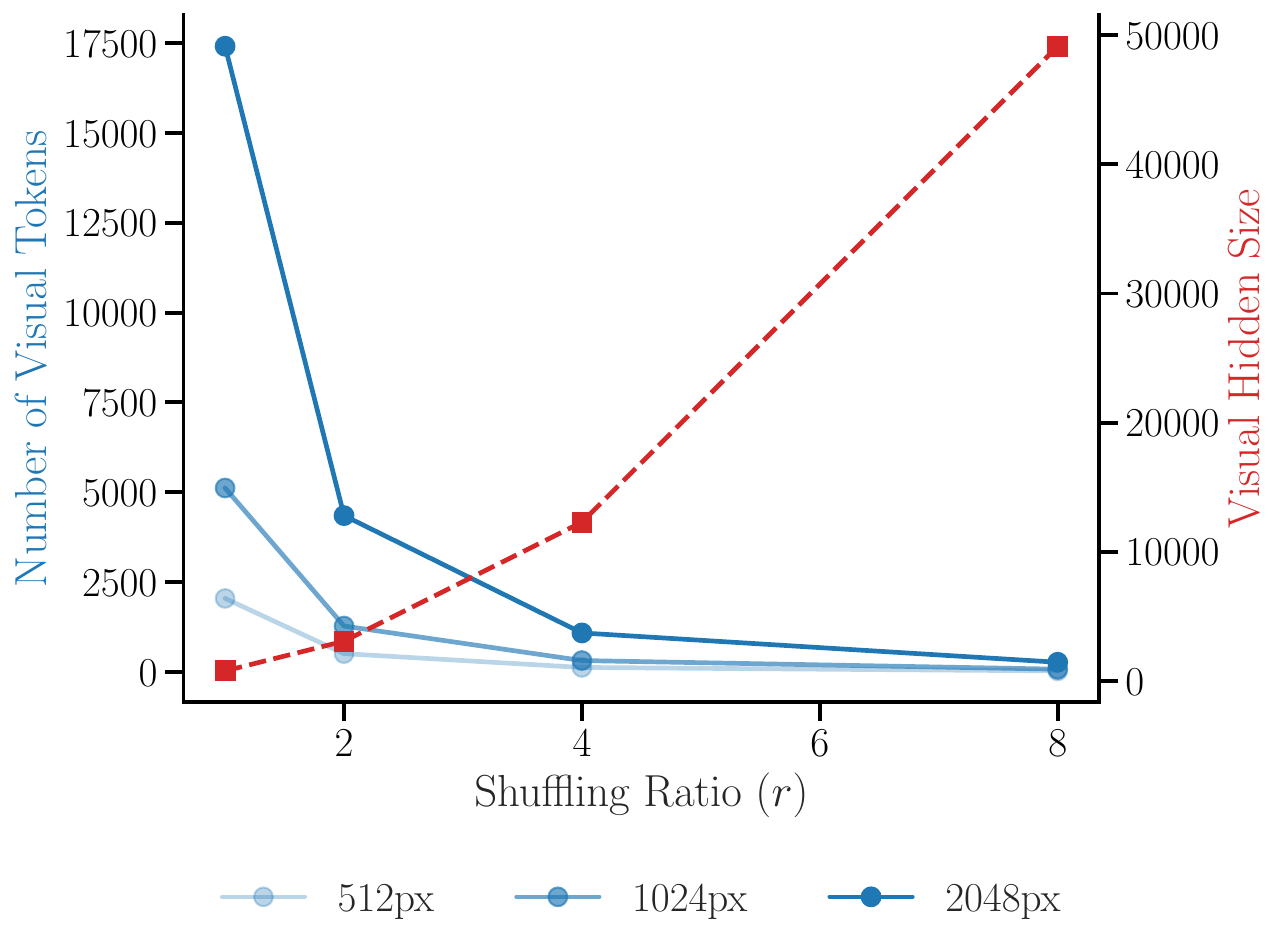} 
\caption{\textbf{\textcolor{red}{Image processing parameters impact on visual tokens.}} Here we assume a square image for simplicity. Scaling the image size introduces naturally more tokens, but having a large enough pixel shuffling ratio ($r\geq4$) allows to counterbalance by concatenating spatially close patch representations.} \label{fig:tokens_shuffling_tradeoff} \end{figure}

\textcolor{red}{Figure~\ref{fig:tokens_shuffling_tradeoff} presents the pixel shuffling trade-off. Processing larger images creates more visual tokens, leading to very long sequences (around $17'500$ tokens for a 2048x2048 px image with no pixel shuffling). Pixel shuffling allow to compress these sequence by concatenating the embeddings of spatially close patches. This diminishes the number of tokens for longer visual token embeddings. Table~\ref{tab:image_resolution_latency} presents the latency to process one image of various resolutions on one L4 GPU and CPU.}

\begin{table}[ht]
    \centering
    \renewcommand{\arraystretch}{1.0}
    \small
    \resizebox{0.85\textwidth}{!}{%
    \begin{tabular}{@{} l c cc}
    & {\textbf{Num. Visual Tokens}} & {\textbf{CPU Latency (ms)}} & {\textbf{GPU Latency (ms)}} \\
    
    \midrule
    $512$px & $128$ & $287.2_{(\pm 7.8)}$ & $43.6 _{(\pm 1.4)}$ \\
    $1024$px & $320$ & $1015.8 _{(\pm 58.1)}$ & $150.3 _{(\pm 2.5)}$ \\
    $2048$px & $1088$ & $2572.0 _{(\pm 63.9)}$ & $363.4 _{(\pm 4.6)}$ \\
    
    \bottomrule
    \end{tabular}%
    }
    \caption{\textcolor{red}{\textbf{\flagship{} image processing latency}}. Computing the average time to process a single image on GPU and CPU. The average is computed on 100 images. The values represent the mean latency in milliseconds, with the standard deviation included in parenthesis.}
    \label{tab:image_resolution_latency}
\end{table}

\subsubsection{Online Query Encoding Latency}
\label{appendix:gpu_latency}
\textcolor{red}{We evaluate the query embedding speed of our model on GPU. We use a single Nvidia H100 with 80GB of VRAM. As for Section~\ref{sec:results}, latencies are computed in batch size 1 to simulate online situations, and are averaged over all NanoBEIR queries. Only the text parameters are loaded and run, to minimize memory usage. Parameters are cast to bfloat16 and Flash Attention 2 is used. The resulting speeds are often much faster than those obtained by running inference through each model's reference implementation. Results are shown in Table~\ref{table:pu_latency}). Interestingly in this setup where memory is not a bottleneck, model depth seems to be a large performance driver, sometimes more the parameter count.
We finally evaluate batched GPU throughput. We use batches of size 512 by default and iteratively half it when memory is insufficient. We observe that \flagship{} based models are extremely fast and can process 5000 queries per second. In the table, the reported figures correspond to the inverted throughput (latency per batch divided by the number of queries per batch). These speed and throughput gains are made possible due to a combination of size, and efficient hardware-informed design as well as the support of flash attention and sequence packing other models of the size often lack \citep{warner2024smarterbetterfasterlonger}.
}

\input{tables/latency_ldb.tex}

\clearpage
\input{tables/natcap_prompt}

%% file: tables/implementation_details.tex
\begin{table*}[ht]
\centering
\resizebox{\textwidth}{!}{%
\begin{tabular}{l r r r r}
  \toprule
  Model & Batch Size & Learning Rate & Training Steps & Training GPU Hours \\
  \midrule
  \multicolumn{5}{l}{\textbf{Modality Alignment}} \\
  \flagshipbase{} (Table~\ref{tab:mod_alignment_mix}) & 4096 & 1e-4 & 5500 & 1920h \\
  \midrule
  \multicolumn{5}{l}{\textbf{Contrastive Learning}} \\
  Generalist contrastive training (Table~\ref{tab:cl_mix}) & 256 & 2e-4 & 3917 & 80h \\
  \midrule
  \multicolumn{5}{l}{\textbf{Document Specialization}} \\
  Document-focused contrastive training w/ hard negatives (Table~\ref{tab:cl_mix}) & 64 & 2e-4 & 19602 & 160h \\
  \bottomrule
\end{tabular}
}
\caption{Training details of our final models at each training stage. GPU Hours are on 80GB H100 GPUs.}
\label{table:app_training_details}
\end{table*}

%% file: tables/alignment_mix.tex
\begin{table}[ht]
\centering
\sisetup{group-separator={,}}
\begin{tabular}{
    l
    S[table-format=7.0]
    S[table-format=8.0]
    S[table-format=10.0]
    S[table-format=3.2]
}
\toprule
\textbf{Dataset Subsection} & {\textbf{\makecell{\# Images}}} & {\textbf{\makecell{\# QA Pairs}}} & {\textbf{\makecell{\# Tokens}}} & {\textbf{\makecell{\% Mix}}} \\
\midrule
Captioning & 609843 & 612768 & 62906011 & 3.13 \\
Real-world VQA & 457360 & 2125615 & 23318335 & 1.16 \\
\makecell[l]{OCR, Document Understanding} & 2499258 & 11415478 & 426806479 & 21.21 \\
Chart/Figure Understanding & 539743 & 24444120 & 30315784 & 1.51 \\
Table Understanding & 163568 & 229077 & 21371931 & 1.06 \\
\makecell[l]{Reasoning, Logic, Maths} & 490870 & 2212629 & 32450213 & 1.61 \\
Screenshot to Code & 547974 & 548296 & 336299551 & 16.71 \\
\makecell[l]{Text-only Instructions} & 0 & 21482682 & 1079001075 & 53.61 \\
\midrule
\textbf{Total} & \textbf{5308616} & \textbf{63070665} & \textbf{2012469379} & \textbf{100.00} \\
\bottomrule
\end{tabular}
\caption{Aggregated statistics of modality alignment datasets from The Cauldron 2~\citep{laurençon2024mattersbuildingvisionlanguagemodels} and Docmatix~\citep{laurençon2024building}, showing image counts, QA pairs, token counts, and the proportional contribution of each subsection to the overall mixture.}
\label{tab:mod_alignment_mix}
\end{table}

%% file: tables/natcap.tex
\begin{table*}[ht]
\small
\setlength{\tabcolsep}{6pt}
\renewcommand{\arraystretch}{1.15}
\begin{tabularx}{\textwidth}{l X S[table-format=6.0]}
\toprule
\textbf{Dataset} & \textbf{Description} & {\textbf{\# Items}} \\
\midrule
Caltech101      & General objects.                                                                  & 3000   \\
Caltech256      & General objects.                                                                  & 30000  \\
Cars            & Car model classification.                                                         & 8000   \\
Country211      & Country where the picture is taken.                                               & 28000  \\
DTD             & Describable textures (texture attributes).                                        & 4000   \\
EuroSat         & Land use / area zone type.                                                        & 16000  \\
FER2013         & Facial emotion recognition.                                                       & 28000  \\
FGCVAircraft    & Aircraft model recognition.                                                       & 3000   \\
Food101         & Food categories.                                                                  & 75000  \\
OxfordPets      & Dog/cat species.                                                                  & 3000   \\
RESISC45        & Aerial scene / area zone type.                                                    & 18000  \\
SUN397          & General scenes.                                                                   & 109000 \\
VOC2007         & General objects.                                                                  & 8000   \\
\midrule
\textbf{TOTAL}  &                                                                                   & \textbf{333000} \\
\bottomrule
\end{tabularx}
\caption{\textbf{\natcap{} Dataset Composition.} \natcap{} spans 13 different sources covering various images types. The total dataset is composed of 333k pairs}
\label{tab:natcap}
\end{table*}

%% file: tables/cl_mix.tex
\begin{table}[ht]
\centering
\setlength{\tabcolsep}{4pt}
\renewcommand{\arraystretch}{1.15}
\begin{tabular}{l p{7cm} r c}
\toprule
\textbf{Source} & \textbf{Description} & \textbf{Pairs} & \textbf{Epochs} \\
\midrule
\multicolumn{4}{l}{\textbf{Generalist Mix}} \\
ColPali~\citep{faysse2025colpaliefficientdocumentretrieval} & Query–Document images for visual retrieval & 118k & 1 \\
MSCOCO~\citep{lin_microsoft_2014} & Natural images with human-written captions & 118k & 1 \\
\natcap~\textit{(ours, subsampled)} & Diverse images with synthetic captions & 118k & 1 \\
RLHN~\citep{thakur2025rlhn} & Text–text pairs for complex retrieval & 680k & 1 \\
\midrule
\textbf{TOTAL} & & \textbf{1030k} & \\
\midrule
\multicolumn{4}{l}{\textbf{Document-Focused Mix}} \\
ColPali~\citep{faysse2025colpaliefficientdocumentretrieval} & Query–Document images for visual retrieval & 118k & 3 \\
RLHN~\citep{thakur2025rlhn} & Text–text pairs for complex retrieval & 300k & 3 \\
\midrule
\textbf{TOTAL} & & \textbf{1254k} & \\
\bottomrule
\end{tabular}
\caption{\textbf{Data mixes for contrastive trainings.} The \textit{Generalist Mix} spans over 1M diverse pairs, while the \textit{Document-Focused Mix} emphasizes document retrieval with extra ColPali epochs.}
\label{tab:cl_mix}
\end{table}

%% file: tables/latency_ldb.tex
\begin{table*}[t]
  \centering
  \renewcommand{\arraystretch}{1.03}
  \small
  \resizebox{\textwidth}{!}{%
  \begin{tabular}{@{} >{}l c c >{\columncolor{latcpucol}}c >{\columncolor{latgpucol}}c >{\columncolor{latcpucol}}c}
    & {Late Interaction} & {Model Size (B)} &
      \rot{\textbf{CPU Latency (ms)} \hspace{5pt}} & \rot{\textbf{GPU Latency (ms)}} & \rot{\textbf{GPU Batching (ms)}} \\

    \midrule
    \multicolumn{6}{@{}l}{\itshape $\ge$ 1B Parameters}\\

    MoCa-3B &  & 3.75 &
      $158_{(\pm 147)}$ & $26_{(\pm 3)}$ & $4.54$ \\

    VLM2Vec &  & 4.15 &
      $211_{(\pm 253)}$ & $21_{(\pm 3)}$ & $2.82$\\

    GME-Qwen2-7B &  & 8.29 &
      $412_{(\pm 411)}$ & $25_{(\pm 1)}$ & $9.07$\\

    E5-V &  & 8.36 &
      $434_{(\pm 379)}$ & $22_{(\pm 2)}$ & $9.55$\\

    ColPali & \cmark & 2.92 &
      $175_{(\pm 113)}$ & $14_{(\pm 1)}$& $3.07$ \\

    ColQwen2.5 & \cmark & 3.75 &
      $158_{(\pm 147)}$ & $26_{(\pm 2)}$ & $26$\\

    Jina-v4 & \cmark & 3.75 &
      $158_{(\pm 147)}$ & $26_{(\pm 2)}$ & $4.54$\\

    NemoRetriever-3B & \cmark & 4.40 &
      $155_{(\pm 118)}$ & $20_{(\pm 2)}$ & $4.59$ \\

    \midrule
    \multicolumn{5}{@{}l}{\itshape $\le$ 1B Parameters}\\

    Jina CLIP &  & .22 &
      $14_{(\pm 7)}$ & $6_{(\pm 2)}$ & $.69$\\

    BGE Visualized M3 &  & .87 &
      $38_{(\pm 42)}$ & $10_{(\pm 2)}$ & $.77$ \\

    SigLIP2-L-512/16 &  & .88 &
      $25_{(\pm 8)}$ & $6_{(\pm 1)}$ & $.10$\\

    ColFlor & \cmark & .17 &
      $17_{(\pm 9)}$ & $8_{(\pm 2)}$ & $.31$\\

    \biflagship{} (ours) &  & .25 &
      $20_{(\pm 11)}$ & $14_{(\pm 2)}$ & $.20$\\

    \textbf{\colflagship{} (ours)} & \cmark & .25 &
      $20_{(\pm 11)}$ & $14_{(\pm 2)}$ & $.20$\\

    \bottomrule
  \end{tabular}%
  }
  \caption{\textcolor{red}{\textbf{Text query encoding latency.}} The latency is computed both on high-end CPUs (1TB RAM, 128 cores) and GPU (Nvidia H100, 80GB) (mean ± std). Since only 649 queries are used, standard deviations are not reported in GPU batching mode (batches of 512 queries by default), for which we report the inverse throughput (average latency per batch divided by the batch size).}
  \label{table:pu_latency}
\end{table*}

%% file: tables/natcap_prompt.tex
\begin{tcolorbox}[colback=gray!5,colframe=black,title=\natcap{} Annotation Prompt,fonttitle=\bfseries]
You are an image annotator expert.\\\\
You will receive an image along with its classification label and the classification task scope, and your task is to provide contextualized metadata about it.\\\\
The output should be a JSON object with the following metadata fields:
\begin{itemize}
  \item \textbf{caption}: A descriptive caption of the image accounting for its label. This should be a \textbf{unique} and concise sentence that describes the image in detail.
  \item \textbf{class\_tags}: A list of tags that represents the image and can help identify the class. (e.g., for a car image with its model as a class, this could be some specific attribute of the car)
  \item \textbf{other\_tags}: A list of tags that represents the image but can help identify the image among others of the same class. (e.g., for a car image with its model as a class, this could be its color or the background of the image)
  \item \textbf{is\_image\_class\_explicit}: Boolean, could the class be inferred from the image alone? (e.g., the class is a country and you cannot necessarily infer it from the image alone, so this would be \texttt{false})
\end{itemize}

Please ensure that the output is in valid JSON format.\\

\textbf{Example:}\\
You receive an image of what is clearly a car with its model as a class (here Audi TTS coupe 2012) for a car model classification task.\\
The output could be a JSON object like this:

\begin{lstlisting}[basicstyle=\ttfamily\small]
{
  "caption": "A red Audi TTS coupe 2012 car parked on a sunny street 
                in front of a sport shop.",
  "class_tags": ["sport coupe","four door coupe","17'' alloy wheels"],
  "other_tags": ["sunny street","parked","red","sport shop"],
  "is_image_class_explicit": true
}
\end{lstlisting}

\textbf{------}\\
Classification scope: \{task\_info\}\\
Image label: \{label\}\\
Answer:
\label{box:natcap_prompt}
\end{tcolorbox}

%% file: iclr2026_conference.bib
@misc{faysse2025colpaliefficientdocumentretrieval,
      title={ColPali: Efficient Document Retrieval with Vision Language Models}, 
      author={Manuel Faysse and Hugues Sibille and Tony Wu and Bilel Omrani and Gautier Viaud and Céline Hudelot and Pierre Colombo},
      year={2025},
      eprint={2407.01449},
      archivePrefix={arXiv},
      primaryClass={cs.IR},
      url={https://arxiv.org/abs/2407.01449}, 
}

@misc{marafioti2025smolvlmredefiningsmallefficient,
      title={SmolVLM: Redefining small and efficient multimodal models}, 
      author={Andrés Marafioti and Orr Zohar and Miquel Farré and Merve Noyan and Elie Bakouch and Pedro Cuenca and Cyril Zakka and Loubna Ben Allal and Anton Lozhkov and Nouamane Tazi and Vaibhav Srivastav and Joshua Lochner and Hugo Larcher and Mathieu Morlon and Lewis Tunstall and Leandro von Werra and Thomas Wolf},
      year={2025},
      eprint={2504.05299},
      archivePrefix={arXiv},
      primaryClass={cs.AI},
      url={https://arxiv.org/abs/2504.05299}, 
}

@misc{laurençon2024mattersbuildingvisionlanguagemodels,
      title={What matters when building vision-language models?}, 
      author={Hugo Laurençon and Léo Tronchon and Matthieu Cord and Victor Sanh},
      year={2024},
      eprint={2405.02246},
      archivePrefix={arXiv},
      primaryClass={cs.CV},
      url={https://arxiv.org/abs/2405.02246}, 
}

@misc{tschannen2025siglip2multilingualvisionlanguage,
      title={SigLIP 2: Multilingual Vision-Language Encoders with Improved Semantic Understanding, Localization, and Dense Features}, 
      author={Michael Tschannen and Alexey Gritsenko and Xiao Wang and Muhammad Ferjad Naeem and Ibrahim Alabdulmohsin and Nikhil Parthasarathy and Talfan Evans and Lucas Beyer and Ye Xia and Basil Mustafa and Olivier Hénaff and Jeremiah Harmsen and Andreas Steiner and Xiaohua Zhai},
      year={2025},
      eprint={2502.14786},
      archivePrefix={arXiv},
      primaryClass={cs.CV},
      url={https://arxiv.org/abs/2502.14786}, 
}

@misc{shi2016realtimesingleimagevideo,
      title={Real-Time Single Image and Video Super-Resolution Using an Efficient Sub-Pixel Convolutional Neural Network}, 
      author={Wenzhe Shi and Jose Caballero and Ferenc Huszár and Johannes Totz and Andrew P. Aitken and Rob Bishop and Daniel Rueckert and Zehan Wang},
      year={2016},
      eprint={1609.05158},
      archivePrefix={arXiv},
      primaryClass={cs.CV},
      url={https://arxiv.org/abs/1609.05158}, 
}

@misc{li2022blipbootstrappinglanguageimagepretraining,
      title={BLIP: Bootstrapping Language-Image Pre-training for Unified Vision-Language Understanding and Generation}, 
      author={Junnan Li and Dongxu Li and Caiming Xiong and Steven Hoi},
      year={2022},
      eprint={2201.12086},
      archivePrefix={arXiv},
      primaryClass={cs.CV},
      url={https://arxiv.org/abs/2201.12086}, 
}

@misc{allal2025smollm2smolgoesbig,
      title={SmolLM2: When Smol Goes Big -- Data-Centric Training of a Small Language Model}, 
      author={Loubna Ben Allal and Anton Lozhkov and Elie Bakouch and Gabriel Martín Blázquez and Guilherme Penedo and Lewis Tunstall and Andrés Marafioti and Hynek Kydlíček and Agustín Piqueres Lajarín and Vaibhav Srivastav and Joshua Lochner and Caleb Fahlgren and Xuan-Son Nguyen and Clémentine Fourrier and Ben Burtenshaw and Hugo Larcher and Haojun Zhao and Cyril Zakka and Mathieu Morlon and Colin Raffel and Leandro von Werra and Thomas Wolf},
      year={2025},
      eprint={2502.02737},
      archivePrefix={arXiv},
      primaryClass={cs.CL},
      url={https://arxiv.org/abs/2502.02737}, 
}

@misc{hu2024minicpmunveilingpotentialsmall,
      title={MiniCPM: Unveiling the Potential of Small Language Models with Scalable Training Strategies}, 
      author={Shengding Hu and Yuge Tu and Xu Han and Chaoqun He and Ganqu Cui and Xiang Long and Zhi Zheng and Yewei Fang and Yuxiang Huang and Weilin Zhao and Xinrong Zhang and Zheng Leng Thai and Kaihuo Zhang and Chongyi Wang and Yuan Yao and Chenyang Zhao and Jie Zhou and Jie Cai and Zhongwu Zhai and Ning Ding and Chao Jia and Guoyang Zeng and Dahai Li and Zhiyuan Liu and Maosong Sun},
      year={2024},
      eprint={2404.06395},
      archivePrefix={arXiv},
      primaryClass={cs.CL},
      url={https://arxiv.org/abs/2404.06395}, 
}

@article{khattab_colbert_2020,
	title = {{ColBERT}: {Efficient} and {Effective} {Passage} {Search} via {Contextualized} {Late} {Interaction} over {BERT}},
	copyright = {arXiv.org perpetual, non-exclusive license},
	shorttitle = {{ColBERT}},
	url = {https://arxiv.org/abs/2004.12832},
	doi = {10.48550/ARXIV.2004.12832},
	urldate = {2024-05-18},
	author = {Khattab, Omar and Zaharia, Matei},
	year = {2020},
}

@article{dosovitskiy_image_2020,
	title = {An {Image} is {Worth} 16x16 {Words}: {Transformers} for {Image} {Recognition} at {Scale}},
	copyright = {arXiv.org perpetual, non-exclusive license},
	shorttitle = {An {Image} is {Worth} 16x16 {Words}},
	url = {https://arxiv.org/abs/2010.11929},
	doi = {10.48550/ARXIV.2010.11929},
	urldate = {2023-10-28},
	author = {Dosovitskiy, Alexey and Beyer, Lucas and Kolesnikov, Alexander and Weissenborn, Dirk and Zhai, Xiaohua and Unterthiner, Thomas and Dehghani, Mostafa and Minderer, Matthias and Heigold, Georg and Gelly, Sylvain and Uszkoreit, Jakob and Houlsby, Neil},
	year = {2020},
	note = {Publisher: arXiv
Version Number: 2},
}

@misc{abdin_phi-3_2024,
	title = {Phi-3 {Technical} {Report}: {A} {Highly} {Capable} {Language} {Model} {Locally} on {Your} {Phone}},
	copyright = {Creative Commons Attribution 4.0 International},
	shorttitle = {Phi-3 {Technical} {Report}},
	url = {https://arxiv.org/abs/2404.14219},
	doi = {10.48550/ARXIV.2404.14219},
	urldate = {2024-05-21},
	publisher = {arXiv},
	author = {Abdin, Marah and Jacobs, Sam Ade and Awan, Ammar Ahmad and Aneja, Jyoti and Awadallah, Ahmed and Awadalla, Hany and Bach, Nguyen and Bahree, Amit and Bakhtiari, Arash and Behl, Harkirat and Benhaim, Alon and Bilenko, Misha and Bjorck, Johan and Bubeck, Sébastien and Cai, Martin and Mendes, Caio César Teodoro and Chen, Weizhu and Chaudhary, Vishrav and Chopra, Parul and Del Giorno, Allie and de Rosa, Gustavo and Dixon, Matthew and Eldan, Ronen and Iter, Dan and Garg, Amit and Goswami, Abhishek and Gunasekar, Suriya and Haider, Emman and Hao, Junheng and Hewett, Russell J. and Huynh, Jamie and Javaheripi, Mojan and Jin, Xin and Kauffmann, Piero and Karampatziakis, Nikos and Kim, Dongwoo and Khademi, Mahoud and Kurilenko, Lev and Lee, James R. and Lee, Yin Tat and Li, Yuanzhi and Liang, Chen and Liu, Weishung and Lin, Eric and Lin, Zeqi and Madan, Piyush and Mitra, Arindam and Modi, Hardik and Nguyen, Anh and Norick, Brandon and Patra, Barun and Perez-Becker, Daniel and Portet, Thomas and Pryzant, Reid and Qin, Heyang and Radmilac, Marko and Rosset, Corby and Roy, Sambudha and Ruwase, Olatunji and Saarikivi, Olli and Saied, Amin and Salim, Adil and Santacroce, Michael and Shah, Shital and Shang, Ning and Sharma, Hiteshi and Song, Xia and Tanaka, Masahiro and Wang, Xin and Ward, Rachel and Wang, Guanhua and Witte, Philipp and Wyatt, Michael and Xu, Can and Xu, Jiahang and Yadav, Sonali and Yang, Fan and Yang, Ziyi and Yu, Donghan and Zhang, Chengruidong and Zhang, Cyril and Zhang, Jianwen and Zhang, Li Lyna and Zhang, Yi and Zhang, Yue and Zhang, Yunan and Zhou, Xiren},
	year = {2024},
	note = {Version Number: 2},
}

@misc{koukounas_jina_2024,
	title = {Jina {CLIP}: {Your} {CLIP} {Model} {Is} {Also} {Your} {Text} {Retriever}},
	copyright = {Creative Commons Attribution 4.0 International},
	shorttitle = {Jina {CLIP}},
	url = {https://arxiv.org/abs/2405.20204},
	doi = {10.48550/ARXIV.2405.20204},
	urldate = {2024-06-01},
	publisher = {arXiv},
	author = {Koukounas, Andreas and Mastrapas, Georgios and Günther, Michael and Wang, Bo and Martens, Scott and Mohr, Isabelle and Sturua, Saba and Akram, Mohammad Kalim and Martínez, Joan Fontanals and Ognawala, Saahil and Guzman, Susana and Werk, Maximilian and Wang, Nan and Xiao, Han},
	year = {2024},
	note = {Version Number: 1},
}

@misc{muennighoff_mteb_2022,
	title = {{MTEB}: {Massive} {Text} {Embedding} {Benchmark}},
	copyright = {arXiv.org perpetual, non-exclusive license},
	shorttitle = {{MTEB}},
	url = {https://arxiv.org/abs/2210.07316},
	doi = {10.48550/ARXIV.2210.07316},
	urldate = {2024-06-02},
	publisher = {arXiv},
	author = {Muennighoff, Niklas and Tazi, Nouamane and Magne, Loïc and Reimers, Nils},
	year = {2022},
	note = {Version Number: 3},
}

@misc{reimers_sentence-bert_2019,
	title = {Sentence-{BERT}: {Sentence} {Embeddings} using {Siamese} {BERT}-{Networks}},
	copyright = {Creative Commons Attribution Share Alike 4.0 International},
	shorttitle = {Sentence-{BERT}},
	url = {https://arxiv.org/abs/1908.10084},
	doi = {10.48550/ARXIV.1908.10084},
	urldate = {2024-06-02},
	publisher = {arXiv},
	author = {Reimers, Nils and Gurevych, Iryna},
	year = {2019},
	note = {Version Number: 1},
}

@misc{liu_improved_2023,
	title = {Improved {Baselines} with {Visual} {Instruction} {Tuning}},
	copyright = {Creative Commons Attribution 4.0 International},
	url = {https://arxiv.org/abs/2310.03744},
	doi = {10.48550/ARXIV.2310.03744},
	urldate = {2024-06-02},
	publisher = {arXiv},
	author = {Liu, Haotian and Li, Chunyuan and Li, Yuheng and Lee, Yong Jae},
	year = {2023},
	note = {Version Number: 2},
}

@article{bai_qwen-vl_2023,
	title = {Qwen-{VL}: {A} {Versatile} {Vision}-{Language} {Model} for {Understanding}, {Localization}, {Text} {Reading}, and {Beyond}},
	copyright = {arXiv.org perpetual, non-exclusive license},
	shorttitle = {Qwen-{VL}},
	url = {https://arxiv.org/abs/2308.12966},
	doi = {10.48550/ARXIV.2308.12966},
	urldate = {2024-02-08},
	author = {Bai, Jinze and Bai, Shuai and Yang, Shusheng and Wang, Shijie and Tan, Sinan and Wang, Peng and Lin, Junyang and Zhou, Chang and Zhou, Jingren},
	year = {2023},
	note = {Publisher: arXiv
Version Number: 3},
}

@misc{laurencon_what_2024,
	title = {What matters when building vision-language models?},
	url = {http://arxiv.org/abs/2405.02246},
	urldate = {2024-05-06},
	publisher = {arXiv},
	author = {Laurençon, Hugo and Tronchon, Léo and Cord, Matthieu and Sanh, Victor},
	month = may,
	year = {2024},
	note = {arXiv:2405.02246 [cs]},
}

@misc{chen_bge_2024,
	title = {{BGE} {M3}-{Embedding}: {Multi}-{Lingual}, {Multi}-{Functionality}, {Multi}-{Granularity} {Text} {Embeddings} {Through} {Self}-{Knowledge} {Distillation}},
	copyright = {Creative Commons Attribution 4.0 International},
	shorttitle = {{BGE} {M3}-{Embedding}},
	url = {https://arxiv.org/abs/2402.03216},
	doi = {10.48550/ARXIV.2402.03216},
	urldate = {2024-06-04},
	publisher = {arXiv},
	author = {Chen, Jianlv and Xiao, Shitao and Zhang, Peitian and Luo, Kun and Lian, Defu and Liu, Zheng},
	year = {2024},
	note = {Version Number: 3},
}

@misc{wang_text_2022,
	title = {Text {Embeddings} by {Weakly}-{Supervised} {Contrastive} {Pre}-training},
	copyright = {Creative Commons Attribution 4.0 International},
	url = {https://arxiv.org/abs/2212.03533},
	doi = {10.48550/ARXIV.2212.03533},
	urldate = {2024-06-04},
	publisher = {arXiv},
	author = {Wang, Liang and Yang, Nan and Huang, Xiaolong and Jiao, Binxing and Yang, Linjun and Jiang, Daxin and Majumder, Rangan and Wei, Furu},
	year = {2022},
	note = {Version Number: 2},
}

@article{faysse_croissantllm_2024,
	title = {{CroissantLLM}: {A} {Truly} {Bilingual} {French}-{English} {Language} {Model}},
	copyright = {arXiv.org perpetual, non-exclusive license},
	shorttitle = {{CroissantLLM}},
	url = {https://arxiv.org/abs/2402.00786},
	doi = {10.48550/ARXIV.2402.00786},
	urldate = {2024-02-18},
	author = {Faysse, Manuel and Fernandes, Patrick and Guerreiro, Nuno M. and Loison, António and Alves, Duarte M. and Corro, Caio and Boizard, Nicolas and Alves, João and Rei, Ricardo and Martins, Pedro H. and Casademunt, Antoni Bigata and Yvon, François and Martins, André F. T. and Viaud, Gautier and Hudelot, Céline and Colombo, Pierre},
	year = {2024},
	note = {Publisher: arXiv
Version Number: 3},
}

@article{touvron_llama_2023,
	title = {Llama 2: {Open} {Foundation} and {Fine}-{Tuned} {Chat} {Models}},
	copyright = {arXiv.org perpetual, non-exclusive license},
	shorttitle = {Llama 2},
	url = {https://arxiv.org/abs/2307.09288},
	doi = {10.48550/ARXIV.2307.09288},
	urldate = {2023-09-03},
	author = {Touvron, Hugo and Martin, Louis and Stone, Kevin and Albert, Peter and Almahairi, Amjad and Babaei, Yasmine and Bashlykov, Nikolay and Batra, Soumya and Bhargava, Prajjwal and Bhosale, Shruti and Bikel, Dan and Blecher, Lukas and Ferrer, Cristian Canton and Chen, Moya and Cucurull, Guillem and Esiobu, David and Fernandes, Jude and Fu, Jeremy and Fu, Wenyin and Fuller, Brian and Gao, Cynthia and Goswami, Vedanuj and Goyal, Naman and Hartshorn, Anthony and Hosseini, Saghar and Hou, Rui and Inan, Hakan and Kardas, Marcin and Kerkez, Viktor and Khabsa, Madian and Kloumann, Isabel and Korenev, Artem and Koura, Punit Singh and Lachaux, Marie-Anne and Lavril, Thibaut and Lee, Jenya and Liskovich, Diana and Lu, Yinghai and Mao, Yuning and Martinet, Xavier and Mihaylov, Todor and Mishra, Pushkar and Molybog, Igor and Nie, Yixin and Poulton, Andrew and Reizenstein, Jeremy and Rungta, Rashi and Saladi, Kalyan and Schelten, Alan and Silva, Ruan and Smith, Eric Michael and Subramanian, Ranjan and Tan, Xiaoqing Ellen and Tang, Binh and Taylor, Ross and Williams, Adina and Kuan, Jian Xiang and Xu, Puxin and Yan, Zheng and Zarov, Iliyan and Zhang, Yuchen and Fan, Angela and Kambadur, Melanie and Narang, Sharan and Rodriguez, Aurelien and Stojnic, Robert and Edunov, Sergey and Scialom, Thomas},
	year = {2023},
	note = {Publisher: arXiv
Version Number: 2},
}

@misc{lucas_beyer_paligemma_2024,
	title = {{PaliGemma}},
	url = {https://www.kaggle.com/m/23393},
	doi = {10.34740/KAGGLE/M/23393},
	urldate = {2024-06-04},
	publisher = {Kaggle},
	author = {{Lucas Beyer*} and {Andreas Steiner*} and {André Susano Pinto*} and {Alexander Kolesnikov*} and {Xiao Wang*} and {Xiaohua Zhai*} and {Daniel Salz} and {Maxim Neumann} and {Ibrahim Alabdulmohsin} and {Michael Tschannen} and {Jeremiah Harmsen} and {Daniel Keysers} and {Neil Houlsby} and {Xi Chen} and {Emanuele Bugliarello} and {Thomas Unterthiner} and {Keran Rong} and {Matthias Minderer} and {Ioana Bica} and {Ivana Balazevic} and {Joan Puigcerver} and {Julian Eisenschlos} and {Manoj Kumar} and {Matko Bošnjak} and {Matthias Bauer} and {Fangyu Liu} and {Adam Grycner} and {Alexey Gritsenko} and {Paul Voigtlaender} and {Pinelopi Papalampidi} and {Olivier Henaff} and {Skanda Koppula} and {Xi Xiong} and {Radu Soricut} and {Model release contributors and general support} and {Tris Warkentin} and {Kat Black} and {Luiz Gustavo Martins} and {Glenn Cameron} and {Raj Gundluru} and {Manvinder Singh} and {Meg Risdal} and {Nilay Chauhan} and {Nate Keating} and {Nesh Devanathan} and {Elisa Bandy} and {Joe Fernandez} and {Antonia Paterson} and {Jenny Brennan} and {Tom Eccles} and {Pankil Botadra} and {Ben Bariach} and {Lav Rai} and {Minwoo Park} and {Dustin Luong} and {Daniel Vlasic} and {Bo Wu} and {Wenming Ye} and {Divyashree Sreepathihalli} and {Kiranbir Sodhia} and {Alek Andreev} and {Armand Joulin} and {Surya Bhupatiraju} and {Minh Giang} and {Joelle Barral} and {Zoubin Ghahramani}},
	year = {2024},
}

@misc{karpukhin_dense_2020,
	title = {Dense {Passage} {Retrieval} for {Open}-{Domain} {Question} {Answering}},
	copyright = {arXiv.org perpetual, non-exclusive license},
	url = {https://arxiv.org/abs/2004.04906},
	doi = {10.48550/ARXIV.2004.04906},
	urldate = {2024-06-14},
	publisher = {arXiv},
	author = {Karpukhin, Vladimir and Oğuz, Barlas and Min, Sewon and Lewis, Patrick and Wu, Ledell and Edunov, Sergey and Chen, Danqi and Yih, Wen-tau},
	year = {2020},
	note = {Version Number: 3},
}

@misc{lewis_retrieval-augmented_2020,
	title = {Retrieval-{Augmented} {Generation} for {Knowledge}-{Intensive} {NLP} {Tasks}},
	copyright = {arXiv.org perpetual, non-exclusive license},
	url = {https://arxiv.org/abs/2005.11401},
	doi = {10.48550/ARXIV.2005.11401},
	urldate = {2024-06-14},
	publisher = {arXiv},
	author = {Lewis, Patrick and Perez, Ethan and Piktus, Aleksandra and Petroni, Fabio and Karpukhin, Vladimir and Goyal, Naman and Küttler, Heinrich and Lewis, Mike and Yih, Wen-tau and Rocktäschel, Tim and Riedel, Sebastian and Kiela, Douwe},
	year = {2020},
	note = {Version Number: 4},
}

@misc{lin_microsoft_2014,
	title = {Microsoft {COCO}: {Common} {Objects} in {Context}},
	copyright = {arXiv.org perpetual, non-exclusive license},
	shorttitle = {Microsoft {COCO}},
	url = {https://arxiv.org/abs/1405.0312},
	doi = {10.48550/ARXIV.1405.0312},
	urldate = {2024-06-14},
	publisher = {arXiv},
	author = {Lin, Tsung-Yi and Maire, Michael and Belongie, Serge and Bourdev, Lubomir and Girshick, Ross and Hays, James and Perona, Pietro and Ramanan, Deva and Zitnick, C. Lawrence and Dollár, Piotr},
	year = {2014},
	note = {Version Number: 3},
}

@misc{cohere_introducing_2024,
	title = {Introducing {Rerank} 3: {A} {New} {Foundation} {Model} for {Efficient} {Enterprise} {Search} \& {Retrieval}},
	url = {https://cohere.com/blog/rerank-3},
	journal = {Cohere Blog},
	author = {Cohere},
	month = apr,
	year = {2024},
}

@misc{beyer2024paligemmaversatile3bvlm,
      title={PaliGemma: A versatile 3B VLM for transfer}, 
      author={Lucas Beyer and Andreas Steiner and André Susano Pinto and Alexander Kolesnikov and Xiao Wang and Daniel Salz and Maxim Neumann and Ibrahim Alabdulmohsin and Michael Tschannen and Emanuele Bugliarello and Thomas Unterthiner and Daniel Keysers and Skanda Koppula and Fangyu Liu and Adam Grycner and Alexey Gritsenko and Neil Houlsby and Manoj Kumar and Keran Rong and Julian Eisenschlos and Rishabh Kabra and Matthias Bauer and Matko Bošnjak and Xi Chen and Matthias Minderer and Paul Voigtlaender and Ioana Bica and Ivana Balazevic and Joan Puigcerver and Pinelopi Papalampidi and Olivier Henaff and Xi Xiong and Radu Soricut and Jeremiah Harmsen and Xiaohua Zhai},
      year={2024},
      eprint={2407.07726},
      archivePrefix={arXiv},
      primaryClass={cs.CV},
      url={https://arxiv.org/abs/2407.07726}, 
}

@misc{wang2024qwen2vlenhancingvisionlanguagemodels,
      title={Qwen2-VL: Enhancing Vision-Language Model's Perception of the World at Any Resolution}, 
      author={Peng Wang and Shuai Bai and Sinan Tan and Shijie Wang and Zhihao Fan and Jinze Bai and Keqin Chen and Xuejing Liu and Jialin Wang and Wenbin Ge and Yang Fan and Kai Dang and Mengfei Du and Xuancheng Ren and Rui Men and Dayiheng Liu and Chang Zhou and Jingren Zhou and Junyang Lin},
      year={2024},
      eprint={2409.12191},
      archivePrefix={arXiv},
      primaryClass={cs.CV},
      url={https://arxiv.org/abs/2409.12191}, 
}

@misc{laurençon2024building,
      title={Building and better understanding vision-language models: insights and future directions.}, 
      author={Hugo Laurençon and Andrés Marafioti and Victor Sanh and Léo Tronchon},
      year={2024},
      eprint={2408.12637},
      archivePrefix={arXiv},
      primaryClass={cs.CV}
}

@misc{dao2023flashattention2fasterattentionbetter,
      title={FlashAttention-2: Faster Attention with Better Parallelism and Work Partitioning}, 
      author={Tri Dao},
      year={2023},
      eprint={2307.08691},
      archivePrefix={arXiv},
      primaryClass={cs.LG},
      url={https://arxiv.org/abs/2307.08691}, 
}

@misc{modernbert,
      title={Smarter, Better, Faster, Longer: A Modern Bidirectional Encoder for Fast, Memory Efficient, and Long Context Finetuning and Inference}, 
      author={Benjamin Warner and Antoine Chaffin and Benjamin Clavié and Orion Weller and Oskar Hallström and Said Taghadouini and Alexis Gallagher and Raja Biswas and Faisal Ladhak and Tom Aarsen and Nathan Cooper and Griffin Adams and Jeremy Howard and Iacopo Poli},
      year={2024},
      eprint={2412.13663},
      archivePrefix={arXiv},
      primaryClass={cs.CL},
      url={https://arxiv.org/abs/2412.13663}, 
}

@misc{nussbaum2025nomicembedtrainingreproducible,
      title={Nomic Embed: Training a Reproducible Long Context Text Embedder}, 
      author={Zach Nussbaum and John X. Morris and Brandon Duderstadt and Andriy Mulyar},
      year={2025},
      eprint={2402.01613},
      archivePrefix={arXiv},
      primaryClass={cs.CL},
      url={https://arxiv.org/abs/2402.01613}, 
}

@misc{li2023generaltextembeddingsmultistage,
      title={Towards General Text Embeddings with Multi-stage Contrastive Learning}, 
      author={Zehan Li and Xin Zhang and Yanzhao Zhang and Dingkun Long and Pengjun Xie and Meishan Zhang},
      year={2023},
      eprint={2308.03281},
      archivePrefix={arXiv},
      primaryClass={cs.CL},
      url={https://arxiv.org/abs/2308.03281}, 
}

@misc{warner2024smarterbetterfasterlonger,
      title={Smarter, Better, Faster, Longer: A Modern Bidirectional Encoder for Fast, Memory Efficient, and Long Context Finetuning and Inference}, 
      author={Benjamin Warner and Antoine Chaffin and Benjamin Clavié and Orion Weller and Oskar Hallström and Said Taghadouini and Alexis Gallagher and Raja Biswas and Faisal Ladhak and Tom Aarsen and Nathan Cooper and Griffin Adams and Jeremy Howard and Iacopo Poli},
      year={2024},
      eprint={2412.13663},
      archivePrefix={arXiv},
      primaryClass={cs.CL},
      url={https://arxiv.org/abs/2412.13663}, 
}

@misc{GTE-ModernColBERT,
title={GTE-ModernColBERT},
author={Chaffin, Antoine},
url={https://huggingface.co/lightonai/GTE-ModernColBERT-v1},
year={2025}
}

@misc{yang2025qwen251mtechnicalreport,
      title={Qwen2.5-1M Technical Report}, 
      author={An Yang and Bowen Yu and Chengyuan Li and Dayiheng Liu and Fei Huang and Haoyan Huang and Jiandong Jiang and Jianhong Tu and Jianwei Zhang and Jingren Zhou and Junyang Lin and Kai Dang and Kexin Yang and Le Yu and Mei Li and Minmin Sun and Qin Zhu and Rui Men and Tao He and Weijia Xu and Wenbiao Yin and Wenyuan Yu and Xiafei Qiu and Xingzhang Ren and Xinlong Yang and Yong Li and Zhiying Xu and Zipeng Zhang},
      year={2025},
      eprint={2501.15383},
      archivePrefix={arXiv},
      primaryClass={cs.CL},
      url={https://arxiv.org/abs/2501.15383}, 
}

@misc{ma2024unifyingmultimodalretrievaldocument,
      title={Unifying Multimodal Retrieval via Document Screenshot Embedding}, 
      author={Xueguang Ma and Sheng-Chieh Lin and Minghan Li and Wenhu Chen and Jimmy Lin},
      year={2024},
      eprint={2406.11251},
      archivePrefix={arXiv},
      primaryClass={cs.IR},
      url={https://arxiv.org/abs/2406.11251}, 
}

@misc{boizard2025eurobertscalingmultilingualencoders,
      title={EuroBERT: Scaling Multilingual Encoders for European Languages}, 
      author={Nicolas Boizard and Hippolyte Gisserot-Boukhlef and Duarte M. Alves and André Martins and Ayoub Hammal and Caio Corro and Céline Hudelot and Emmanuel Malherbe and Etienne Malaboeuf and Fanny Jourdan and Gabriel Hautreux and João Alves and Kevin El-Haddad and Manuel Faysse and Maxime Peyrard and Nuno M. Guerreiro and Patrick Fernandes and Ricardo Rei and Pierre Colombo},
      year={2025},
      eprint={2503.05500},
      archivePrefix={arXiv},
      primaryClass={cs.CL},
      url={https://arxiv.org/abs/2503.05500}, 
}

@article{macé2025vidorebenchmarkv2raising,
  title={ViDoRe Benchmark V2: Raising the Bar for Visual Retrieval},
  author={Mac{\'e}, Quentin and Loison, Ant{\'o}nio and Faysse, Manuel},
  journal={arXiv preprint arXiv:2505.17166},
  year={2025}
}

@misc{gisserotboukhlef2025pretrainencodersmaskedlanguage,
      title={Should We Still Pretrain Encoders with Masked Language Modeling?}, 
      author={Hippolyte Gisserot-Boukhlef and Nicolas Boizard and Manuel Faysse and Duarte M. Alves and Emmanuel Malherbe and André F. T. Martins and Céline Hudelot and Pierre Colombo},
      year={2025},
      eprint={2507.00994},
      archivePrefix={arXiv},
      primaryClass={cs.CL},
      url={https://arxiv.org/abs/2507.00994}, 
}

@misc{jiang2025vlm2vectrainingvisionlanguagemodels,
      title={VLM2Vec: Training Vision-Language Models for Massive Multimodal Embedding Tasks}, 
      author={Ziyan Jiang and Rui Meng and Xinyi Yang and Semih Yavuz and Yingbo Zhou and Wenhu Chen},
      year={2025},
      eprint={2410.05160},
      archivePrefix={arXiv},
      primaryClass={cs.CV},
      url={https://arxiv.org/abs/2410.05160}, 
}

@misc{chen2025mocamodalityawarecontinualpretraining,
      title={MoCa: Modality-aware Continual Pre-training Makes Better Bidirectional Multimodal Embeddings}, 
      author={Haonan Chen and Hong Liu and Yuping Luo and Liang Wang and Nan Yang and Furu Wei and Zhicheng Dou},
      year={2025},
      eprint={2506.23115},
      archivePrefix={arXiv},
      primaryClass={cs.CV},
      url={https://arxiv.org/abs/2506.23115}, 
}

@misc{jiang2024e5vuniversalembeddingsmultimodal,
      title={E5-V: Universal Embeddings with Multimodal Large Language Models}, 
      author={Ting Jiang and Minghui Song and Zihan Zhang and Haizhen Huang and Weiwei Deng and Feng Sun and Qi Zhang and Deqing Wang and Fuzhen Zhuang},
      year={2024},
      eprint={2407.12580},
      archivePrefix={arXiv},
      primaryClass={cs.CL},
      url={https://arxiv.org/abs/2407.12580}, 
}

@misc{radford2021learningtransferablevisualmodels,
      title={Learning Transferable Visual Models From Natural Language Supervision}, 
      author={Alec Radford and Jong Wook Kim and Chris Hallacy and Aditya Ramesh and Gabriel Goh and Sandhini Agarwal and Girish Sastry and Amanda Askell and Pamela Mishkin and Jack Clark and Gretchen Krueger and Ilya Sutskever},
      year={2021},
      eprint={2103.00020},
      archivePrefix={arXiv},
      primaryClass={cs.CV},
      url={https://arxiv.org/abs/2103.00020}, 
}

@misc{lin2023sphinxjointmixingweights,
      title={SPHINX: The Joint Mixing of Weights, Tasks, and Visual Embeddings for Multi-modal Large Language Models}, 
      author={Ziyi Lin and Chris Liu and Renrui Zhang and Peng Gao and Longtian Qiu and Han Xiao and Han Qiu and Chen Lin and Wenqi Shao and Keqin Chen and Jiaming Han and Siyuan Huang and Yichi Zhang and Xuming He and Hongsheng Li and Yu Qiao},
      year={2023},
      eprint={2311.07575},
      archivePrefix={arXiv},
      primaryClass={cs.CV},
      url={https://arxiv.org/abs/2311.07575}, 
}

@misc{alayrac2022flamingovisuallanguagemodel,
      title={Flamingo: a Visual Language Model for Few-Shot Learning}, 
      author={Jean-Baptiste Alayrac and Jeff Donahue and Pauline Luc and Antoine Miech and Iain Barr and Yana Hasson and Karel Lenc and Arthur Mensch and Katie Millican and Malcolm Reynolds and Roman Ring and Eliza Rutherford and Serkan Cabi and Tengda Han and Zhitao Gong and Sina Samangooei and Marianne Monteiro and Jacob Menick and Sebastian Borgeaud and Andrew Brock and Aida Nematzadeh and Sahand Sharifzadeh and Mikolaj Binkowski and Ricardo Barreira and Oriol Vinyals and Andrew Zisserman and Karen Simonyan},
      year={2022},
      eprint={2204.14198},
      archivePrefix={arXiv},
      primaryClass={cs.CV},
      url={https://arxiv.org/abs/2204.14198}, 
}

@misc{weller2025seqvsseqopen,
      title={Seq vs Seq: An Open Suite of Paired Encoders and Decoders}, 
      author={Orion Weller and Kathryn Ricci and Marc Marone and Antoine Chaffin and Dawn Lawrie and Benjamin Van Durme},
      year={2025},
      eprint={2507.11412},
      archivePrefix={arXiv},
      primaryClass={cs.CL},
      url={https://arxiv.org/abs/2507.11412}, 
}

@misc{mckinzie2024mm1methodsanalysis,
      title={MM1: Methods, Analysis \& Insights from Multimodal LLM Pre-training}, 
      author={Brandon McKinzie and Zhe Gan and Jean-Philippe Fauconnier and Sam Dodge and Bowen Zhang and Philipp Dufter and Dhruti Shah and Xianzhi Du and Futang Peng and Floris Weers and Anton Belyi and Haotian Zhang and Karanjeet Singh and Doug Kang and Ankur Jain and Hongyu Hè and Max Schwarzer and Tom Gunter and Xiang Kong and Aonan Zhang and Jianyu Wang and Chong Wang and Nan Du and Tao Lei and Sam Wiseman and Guoli Yin and Mark Lee and Zirui Wang and Ruoming Pang and Peter Grasch and Alexander Toshev and Yinfei Yang},
      year={2024},
      eprint={2403.09611},
      archivePrefix={arXiv},
      primaryClass={cs.CV},
      url={https://arxiv.org/abs/2403.09611}, 
}

@misc{hu2024mplugdocowl2highresolutioncompressingocrfree,
      title={mPLUG-DocOwl2: High-resolution Compressing for OCR-free Multi-page Document Understanding}, 
      author={Anwen Hu and Haiyang Xu and Liang Zhang and Jiabo Ye and Ming Yan and Ji Zhang and Qin Jin and Fei Huang and Jingren Zhou},
      year={2024},
      eprint={2409.03420},
      archivePrefix={arXiv},
      primaryClass={cs.CV},
      url={https://arxiv.org/abs/2409.03420}, 
}

@misc{xiao2025miebmassiveimageembedding,
      title={MIEB: Massive Image Embedding Benchmark}, 
      author={Chenghao Xiao and Isaac Chung and Imene Kerboua and Jamie Stirling and Xin Zhang and Márton Kardos and Roman Solomatin and Noura Al Moubayed and Kenneth Enevoldsen and Niklas Muennighoff},
      year={2025},
      eprint={2504.10471},
      archivePrefix={arXiv},
      primaryClass={cs.CV},
      url={https://arxiv.org/abs/2504.10471}, 
}

@misc{zhang2025gmeimprovinguniversalmultimodal,
      title={GME: Improving Universal Multimodal Retrieval by Multimodal LLMs}, 
      author={Xin Zhang and Yanzhao Zhang and Wen Xie and Mingxin Li and Ziqi Dai and Dingkun Long and Pengjun Xie and Meishan Zhang and Wenjie Li and Min Zhang},
      year={2025},
      eprint={2412.16855},
      archivePrefix={arXiv},
      primaryClass={cs.CL},
      url={https://arxiv.org/abs/2412.16855}, 
}

@misc{liu2023visualinstructiontuning,
      title={Visual Instruction Tuning}, 
      author={Haotian Liu and Chunyuan Li and Qingyang Wu and Yong Jae Lee},
      year={2023},
      eprint={2304.08485},
      archivePrefix={arXiv},
      primaryClass={cs.CV},
      url={https://arxiv.org/abs/2304.08485}, 
}

@misc{shi2024instructiontuninglossinstructions,
      title={Instruction Tuning With Loss Over Instructions}, 
      author={Zhengyan Shi and Adam X. Yang and Bin Wu and Laurence Aitchison and Emine Yilmaz and Aldo Lipani},
      year={2024},
      eprint={2405.14394},
      archivePrefix={arXiv},
      primaryClass={cs.CL},
      url={https://arxiv.org/abs/2405.14394}, 
}

@misc{huertaenochian2024instructionfinetuningdoesprompt,
      title={Instruction Fine-Tuning: Does Prompt Loss Matter?}, 
      author={Mathew Huerta-Enochian and Seung Yong Ko},
      year={2024},
      eprint={2401.13586},
      archivePrefix={arXiv},
      primaryClass={cs.LG},
      url={https://arxiv.org/abs/2401.13586}, 
}

@misc{hu2021loralowrankadaptationlarge,
      title={LoRA: Low-Rank Adaptation of Large Language Models}, 
      author={Edward J. Hu and Yelong Shen and Phillip Wallis and Zeyuan Allen-Zhu and Yuanzhi Li and Shean Wang and Lu Wang and Weizhu Chen},
      year={2021},
      eprint={2106.09685},
      archivePrefix={arXiv},
      primaryClass={cs.CL},
      url={https://arxiv.org/abs/2106.09685}, 
}

@misc{oord2019representationlearningcontrastivepredictive,
      title={Representation Learning with Contrastive Predictive Coding}, 
      author={Aaron van den Oord and Yazhe Li and Oriol Vinyals},
      year={2019},
      eprint={1807.03748},
      archivePrefix={arXiv},
      primaryClass={cs.LG},
      url={https://arxiv.org/abs/1807.03748}, 
}

@misc{ye2023ureaderuniversalocrfreevisuallysituated,
      title={UReader: Universal OCR-free Visually-situated Language Understanding with Multimodal Large Language Model}, 
      author={Jiabo Ye and Anwen Hu and Haiyang Xu and Qinghao Ye and Ming Yan and Guohai Xu and Chenliang Li and Junfeng Tian and Qi Qian and Ji Zhang and Qin Jin and Liang He and Xin Alex Lin and Fei Huang},
      year={2023},
      eprint={2310.05126},
      archivePrefix={arXiv},
      primaryClass={cs.CV},
      url={https://arxiv.org/abs/2310.05126}, 
}

@misc{zhou2024vistavisualizedtextembedding,
      title={VISTA: Visualized Text Embedding For Universal Multi-Modal Retrieval}, 
      author={Junjie Zhou and Zheng Liu and Shitao Xiao and Bo Zhao and Yongping Xiong},
      year={2024},
      eprint={2406.04292},
      archivePrefix={arXiv},
      primaryClass={cs.IR},
      url={https://arxiv.org/abs/2406.04292}, 
}

@misc{lee2025nvembedimprovedtechniquestraining,
      title={NV-Embed: Improved Techniques for Training LLMs as Generalist Embedding Models}, 
      author={Chankyu Lee and Rajarshi Roy and Mengyao Xu and Jonathan Raiman and Mohammad Shoeybi and Bryan Catanzaro and Wei Ping},
      year={2025},
      eprint={2405.17428},
      archivePrefix={arXiv},
      primaryClass={cs.CL},
      url={https://arxiv.org/abs/2405.17428}, 
}

@misc{günther2025jinaembeddingsv4universalembeddingsmultimodal,
      title={jina-embeddings-v4: Universal Embeddings for Multimodal Multilingual Retrieval}, 
      author={Michael Günther and Saba Sturua and Mohammad Kalim Akram and Isabelle Mohr and Andrei Ungureanu and Bo Wang and Sedigheh Eslami and Scott Martens and Maximilian Werk and Nan Wang and Han Xiao},
      year={2025},
      eprint={2506.18902},
      archivePrefix={arXiv},
      primaryClass={cs.AI},
      url={https://arxiv.org/abs/2506.18902}, 
}

@misc{lin2015microsoftcococommonobjects,
      title={Microsoft COCO: Common Objects in Context}, 
      author={Tsung-Yi Lin and Michael Maire and Serge Belongie and Lubomir Bourdev and Ross Girshick and James Hays and Pietro Perona and Deva Ramanan and C. Lawrence Zitnick and Piotr Dollár},
      year={2015},
      eprint={1405.0312},
      archivePrefix={arXiv},
      primaryClass={cs.CV},
      url={https://arxiv.org/abs/1405.0312}, 
}

@misc{plummer2016flickr30kentitiescollectingregiontophrase,
      title={Flickr30k Entities: Collecting Region-to-Phrase Correspondences for Richer Image-to-Sentence Models}, 
      author={Bryan A. Plummer and Liwei Wang and Chris M. Cervantes and Juan C. Caicedo and Julia Hockenmaier and Svetlana Lazebnik},
      year={2016},
      eprint={1505.04870},
      archivePrefix={arXiv},
      primaryClass={cs.CV},
      url={https://arxiv.org/abs/1505.04870}, 
}

@misc{thakur2025rlhn,
      title={Fixing Data That Hurts Performance: Cascading LLMs to Relabel Hard Negatives for Robust Information Retrieval}, 
      author={Nandan Thakur and Crystina Zhang and Xueguang Ma and Jimmy Lin},
      year={2025},
      eprint={2505.16967},
      archivePrefix={arXiv},
      primaryClass={cs.IR},
      url={https://arxiv.org/abs/2505.16967}, 
}

@misc{xu2025llamanemoretrievercolembedtopperforming,
      title={Llama Nemoretriever Colembed: Top-Performing Text-Image Retrieval Model}, 
      author={Mengyao Xu and Gabriel Moreira and Ronay Ak and Radek Osmulski and Yauhen Babakhin and Zhiding Yu and Benedikt Schifferer and Even Oldridge},
      year={2025},
      eprint={2507.05513},
      archivePrefix={arXiv},
      primaryClass={cs.CV},
      url={https://arxiv.org/abs/2507.05513}, 
}

@misc{wei2022emergentabilitieslargelanguage,
      title={Emergent Abilities of Large Language Models}, 
      author={Jason Wei and Yi Tay and Rishi Bommasani and Colin Raffel and Barret Zoph and Sebastian Borgeaud and Dani Yogatama and Maarten Bosma and Denny Zhou and Donald Metzler and Ed H. Chi and Tatsunori Hashimoto and Oriol Vinyals and Percy Liang and Jeff Dean and William Fedus},
      year={2022},
      eprint={2206.07682},
      archivePrefix={arXiv},
      primaryClass={cs.CL},
      url={https://arxiv.org/abs/2206.07682}, 
}

@misc{clavié2024bettermonolingualjapaneseretrievers,
      title={Towards Better Monolingual Japanese Retrievers with Multi-Vector Models}, 
      author={Benjamin Clavié},
      year={2024},
      eprint={2312.16144},
      archivePrefix={arXiv},
      primaryClass={cs.CL},
      url={https://arxiv.org/abs/2312.16144}, 
}

@misc{allenzhu2025physicslanguagemodels1,
      title={Physics of Language Models: Part 1, Learning Hierarchical Language Structures}, 
      author={Zeyuan Allen-Zhu and Yuanzhi Li},
      year={2025},
      eprint={2305.13673},
      archivePrefix={arXiv},
      primaryClass={cs.CL},
      url={https://arxiv.org/abs/2305.13673}, 
}

@misc{xiao2025metaembedscalingmultimodalretrieval,
      title={MetaEmbed: Scaling Multimodal Retrieval at Test-Time with Flexible Late Interaction}, 
      author={Zilin Xiao and Qi Ma and Mengting Gu and Chun-cheng Jason Chen and Xintao Chen and Vicente Ordonez and Vijai Mohan},
      year={2025},
      eprint={2509.18095},
      archivePrefix={arXiv},
      primaryClass={cs.IR},
      url={https://arxiv.org/abs/2509.18095}, 
}

@article{masrycolflor,
  title={ColFlor: Towards BERT-Size Vision-Language Document Retrieval Models},
  author={Masry, Ahmed and Hoque, Enamul},
  year={2024},

}

@misc{xiao2023florence2advancingunifiedrepresentation,
      title={Florence-2: Advancing a Unified Representation for a Variety of Vision Tasks}, 
      author={Bin Xiao and Haiping Wu and Weijian Xu and Xiyang Dai and Houdong Hu and Yumao Lu and Michael Zeng and Ce Liu and Lu Yuan},
      year={2023},
      eprint={2311.06242},
      archivePrefix={arXiv},
      primaryClass={cs.CV},
      url={https://arxiv.org/abs/2311.06242}, 
}

@misc{zhang2025qwen3embeddingadvancingtext,
      title={Qwen3 Embedding: Advancing Text Embedding and Reranking Through Foundation Models}, 
      author={Yanzhao Zhang and Mingxin Li and Dingkun Long and Xin Zhang and Huan Lin and Baosong Yang and Pengjun Xie and An Yang and Dayiheng Liu and Junyang Lin and Fei Huang and Jingren Zhou},
      year={2025},
      eprint={2506.05176},
      archivePrefix={arXiv},
      primaryClass={cs.CL},
      url={https://arxiv.org/abs/2506.05176}, 
}

@misc{li2024improvinggeneraltextembeddingmerging,
      title={Improving General Text Embedding Model: Tackling Task Conflict and Data Imbalance through Model Merging}, 
      author={Mingxin Li and Zhijie Nie and Yanzhao Zhang and Dingkun Long and Richong Zhang and Pengjun Xie},
      year={2024},
      eprint={2410.15035},
      archivePrefix={arXiv},
      primaryClass={cs.CL},
      url={https://arxiv.org/abs/2410.15035}, 
}

@misc{dziadzio2024mergemultimodalmodelstime,
      title={How to Merge Your Multimodal Models Over Time?}, 
      author={Sebastian Dziadzio and Vishaal Udandarao and Karsten Roth and Ameya Prabhu and Zeynep Akata and Samuel Albanie and Matthias Bethge},
      year={2024},
      eprint={2412.06712},
      archivePrefix={arXiv},
      primaryClass={cs.LG},
      url={https://arxiv.org/abs/2412.06712}, 
}

@misc{sung2023empiricalstudymultimodalmodelmerging,
      title={An Empirical Study of Multimodal Model Merging}, 
      author={Yi-Lin Sung and Linjie Li and Kevin Lin and Zhe Gan and Mohit Bansal and Lijuan Wang},
      year={2023},
      eprint={2304.14933},
      archivePrefix={arXiv},
      primaryClass={cs.CV},
      url={https://arxiv.org/abs/2304.14933}, 
}

@misc{ilharco2022patchingopenvocabularymodelsinterpolating,
      title={Patching open-vocabulary models by interpolating weights}, 
      author={Gabriel Ilharco and Mitchell Wortsman and Samir Yitzhak Gadre and Shuran Song and Hannaneh Hajishirzi and Simon Kornblith and Ali Farhadi and Ludwig Schmidt},
      year={2022},
      eprint={2208.05592},
      archivePrefix={arXiv},
      primaryClass={cs.CV},
      url={https://arxiv.org/abs/2208.05592}, 
}

@article{slerp_original,
author = {Shoemake, Ken},
title = {Animating rotation with quaternion curves},
year = {1985},
issue_date = {Jul. 1985},
publisher = {Association for Computing Machinery},
address = {New York, NY, USA},
volume = {19},
number = {3},
issn = {0097-8930},
url = {https://doi.org/10.1145/325165.325242},
doi = {10.1145/325165.325242},
month = jul,
pages = {245–254},
numpages = {10}
}

@InProceedings{pmlr-v202-fernandes23a,
  title = 	 {Scaling Laws for Multilingual Neural Machine Translation},
  author =       {Fernandes, Patrick and Ghorbani, Behrooz and Garcia, Xavier and Freitag, Markus and Firat, Orhan},
  booktitle = 	 {Proceedings of the 40th International Conference on Machine Learning},
  pages = 	 {10053--10071},
  year = 	 {2023},
  editor = 	 {Krause, Andreas and Brunskill, Emma and Cho, Kyunghyun and Engelhardt, Barbara and Sabato, Sivan and Scarlett, Jonathan},
  volume = 	 {202},
  series = 	 {Proceedings of Machine Learning Research},
  month = 	 {23--29 Jul},
  publisher =    {PMLR},
  url = 	 {https://proceedings.mlr.press/v202/fernandes23a.html}
}

@INPROCEEDINGS{6755945,
  author={Krause, Jonathan and Stark, Michael and Deng, Jia and Fei-Fei, Li},
  booktitle={2013 IEEE International Conference on Computer Vision Workshops}, 
  title={3D Object Representations for Fine-Grained Categorization}, 
  year={2013},
  volume={},
  number={},
  pages={554-561},
  doi={10.1109/ICCVW.2013.77}}

@misc{khaireddin2021facialemotionrecognitionstate,
      title={Facial Emotion Recognition: State of the Art Performance on FER2013}, 
      author={Yousif Khaireddin and Zhuofa Chen},
      year={2021},
      eprint={2105.03588},
      archivePrefix={arXiv},
      primaryClass={cs.CV},
      url={https://arxiv.org/abs/2105.03588}, 
}

@misc{helber2019eurosatnoveldatasetdeep,
      title={EuroSAT: A Novel Dataset and Deep Learning Benchmark for Land Use and Land Cover Classification}, 
      author={Patrick Helber and Benjamin Bischke and Andreas Dengel and Damian Borth},
      year={2019},
      eprint={1709.00029},
      archivePrefix={arXiv},
      primaryClass={cs.CV},
      url={https://arxiv.org/abs/1709.00029}, 
}

@InProceedings{10.1007/978-3-319-10599-4_29,
author="Bossard, Lukas
and Guillaumin, Matthieu
and Van Gool, Luc",
editor="Fleet, David
and Pajdla, Tomas
and Schiele, Bernt
and Tuytelaars, Tinne",
title="Food-101 -- Mining Discriminative Components with Random Forests",
booktitle="Computer Vision -- ECCV 2014",
year="2014",
publisher="Springer International Publishing",
address="Cham",
pages="446--461",
isbn="978-3-319-10599-4"
}

@misc{chen2016trainingdeepnetssublinear,
      title={Training Deep Nets with Sublinear Memory Cost}, 
      author={Tianqi Chen and Bing Xu and Chiyuan Zhang and Carlos Guestrin},
      year={2016},
      eprint={1604.06174},
      archivePrefix={arXiv},
      primaryClass={cs.LG},
      url={https://arxiv.org/abs/1604.06174}, 
}

@misc{rajbhandari2020zeromemoryoptimizationstraining,
      title={ZeRO: Memory Optimizations Toward Training Trillion Parameter Models}, 
      author={Samyam Rajbhandari and Jeff Rasley and Olatunji Ruwase and Yuxiong He},
      year={2020},
      eprint={1910.02054},
      archivePrefix={arXiv},
      primaryClass={cs.LG},
      url={https://arxiv.org/abs/1910.02054}, 
}

@article{Lannelongue2021GreenAlgorithms,
  title     = {Green Algorithms: Quantifying the Carbon Footprint of Computation},
  author    = {Lannelongue, Lo{\"\i}c and Grealey, Joe and Inouye, Michael},
  journal   = {Advances in Science},
  volume    = {7},
  number    = {34},
  pages     = {eabf3899},
  year      = {2021},
  publisher = {American Association for the Advancement of Science},
  doi       = {10.1126/sciadv.abf3899}
}

@inproceedings{Strubell2019Energy,
  title     = {Energy and Policy Considerations for Deep Learning in NLP},
  author    = {Strubell, Emma and Ganesh, Ananya and McCallum, Andrew},
  booktitle = {Proceedings of the 57th Annual Meeting of the Association for Computational Linguistics},
  pages     = {3645--3650},
  year      = {2019},
  publisher = {Association for Computational Linguistics},
  doi       = {10.18653/v1/P19-1355}
}

@article{Patterson2021Carbon,
  title   = {The Carbon Footprint of Machine Learning Training Will Plateau, Then Shrink},
  author  = {Patterson, David and Gonzalez, Joseph and Le, Quoc and Liang, Chen and Munguia, Haisam and Rothchild, Daniel and So, David and Texier, Maud and Dean, Jeff},
  journal = {IEEE Computer},
  volume  = {54},
  number  = {12},
  pages   = {18--28},
  year    = {2021},
  doi     = {10.1109/MC.2021.3120015}
}

@misc{vespaScalingColPali,
	author = {Jo Bergum},
	title = {{S}caling {C}ol{P}ali to billions of {P}{D}{F}s with {V}espa --- blog.vespa.ai},
	howpublished = {\url{https://blog.vespa.ai/scaling-colpali-to-billions/}},
	year = {2025},
}

@misc{enevoldsen2025mmtebmassivemultilingualtext,
      title={MMTEB: Massive Multilingual Text Embedding Benchmark}, 
      author={Kenneth Enevoldsen and Isaac Chung and Imene Kerboua and Márton Kardos and Ashwin Mathur and David Stap and Jay Gala and Wissam Siblini and Dominik Krzemiński and Genta Indra Winata and Saba Sturua and Saiteja Utpala and Mathieu Ciancone and Marion Schaeffer and Gabriel Sequeira and Diganta Misra and Shreeya Dhakal and Jonathan Rystrøm and Roman Solomatin and Ömer Çağatan and Akash Kundu and Martin Bernstorff and Shitao Xiao and Akshita Sukhlecha and Bhavish Pahwa and Rafał Poświata and Kranthi Kiran GV and Shawon Ashraf and Daniel Auras and Björn Plüster and Jan Philipp Harries and Loïc Magne and Isabelle Mohr and Mariya Hendriksen and Dawei Zhu and Hippolyte Gisserot-Boukhlef and Tom Aarsen and Jan Kostkan and Konrad Wojtasik and Taemin Lee and Marek Šuppa and Crystina Zhang and Roberta Rocca and Mohammed Hamdy and Andrianos Michail and John Yang and Manuel Faysse and Aleksei Vatolin and Nandan Thakur and Manan Dey and Dipam Vasani and Pranjal Chitale and Simone Tedeschi and Nguyen Tai and Artem Snegirev and Michael Günther and Mengzhou Xia and Weijia Shi and Xing Han Lù and Jordan Clive and Gayatri Krishnakumar and Anna Maksimova and Silvan Wehrli and Maria Tikhonova and Henil Panchal and Aleksandr Abramov and Malte Ostendorff and Zheng Liu and Simon Clematide and Lester James Miranda and Alena Fenogenova and Guangyu Song and Ruqiya Bin Safi and Wen-Ding Li and Alessia Borghini and Federico Cassano and Hongjin Su and Jimmy Lin and Howard Yen and Lasse Hansen and Sara Hooker and Chenghao Xiao and Vaibhav Adlakha and Orion Weller and Siva Reddy and Niklas Muennighoff},
      year={2025},
      eprint={2502.13595},
      archivePrefix={arXiv},
      primaryClass={cs.CL},
      url={https://arxiv.org/abs/2502.13595}, 
}

@article{conti2025context,
  title={Context is Gold to find the Gold Passage: Evaluating and Training Contextual Document Embeddings},
  author={Conti, Max and Faysse, Manuel and Viaud, Gautier and Bosselut, Antoine and Hudelot, C{\'e}line and Colombo, Pierre},
  journal={arXiv preprint arXiv:2505.24782},
  year={2025}
}

@inproceedings{lin-byrne-2022-retrieval,
    title = "Retrieval Augmented Visual Question Answering with Outside Knowledge",
    author = "Lin, Weizhe  and
      Byrne, Bill",
    editor = "Goldberg, Yoav  and
      Kozareva, Zornitsa  and
      Zhang, Yue",
    booktitle = "Proceedings of the 2022 Conference on Empirical Methods in Natural Language Processing",
    month = dec,
    year = "2022",
    address = "Abu Dhabi, United Arab Emirates",
    publisher = "Association for Computational Linguistics",
    url = "https://aclanthology.org/2022.emnlp-main.772/",
    doi = "10.18653/v1/2022.emnlp-main.772",
    pages = "11238--11254"
}
